\documentclass[twocolumn,secnumarabic,amssymb,nobibnotes,aps,pra,showpacs,superscriptaddress]{revtex4-1}
\usepackage[english]{babel}
\usepackage{graphicx}
\usepackage{amsmath}
\usepackage{amsfonts}
\usepackage[usenames,dvipsnames]{color}
\usepackage{hyperref}
\hypersetup{colorlinks}
\hypersetup{linkcolor=black, citecolor=black, urlcolor=blue}
\usepackage[utf8]{inputenc}

\begin{document}
\title{A Landau-de Gennes theory for hard colloidal rods: defects and tactoids}
\author{J. C. Everts}
\email{j.c.everts@uu.nl}
\address{Institute for Theoretical Physics, Center for Extreme Matter and Emergent Phenomena,  Utrecht University, Princetonplein 5, 3584 CC Utrecht, The Netherlands}
\author{M. T. J. J. M. Punter}
\address{FOM Institute AMOLF, Science Park 104, 1098XG Amsterdam, The Netherlands}
\author{S. Samin}
\address{Institute for Theoretical Physics, Center for Extreme Matter and Emergent Phenomena,  Utrecht University, Princetonplein 5, 3584 CC Utrecht, The Netherlands}
\author{P. van der Schoot}
\address{Institute for Theoretical Physics, Center for Extreme Matter and Emergent Phenomena,  Utrecht University, Princetonplein 5, 3584 CC Utrecht, The Netherlands}
\address{Theory of Polymers and Soft Matter Group, Department of Applied Physics, Eindhoven University of Technology, P.O. Box 513, 5600 MB Eindhoven, The Netherlands}
\author{R. van Roij}
\address{Institute for Theoretical Physics, Center for Extreme Matter and Emergent Phenomena,  Utrecht University, Princetonplein 5, 3584 CC  Utrecht, The Netherlands}
\date{\today}

\begin{abstract}
We construct a phenomenological Landau-de Gennes theory for hard colloidal rods by performing an order parameter expansion of the chemical-potential dependent grand potential. By fitting the coefficients to known results of Onsager theory, we are not only able to describe the isotropic-nematic phase transition as function of density, including the well-known density jump, but also the isotropic-nematic planar interface. The resulting theory is applied in calculations of the isotropic core size in a radial hedgehog defect, the density dependence of linear defects of hard rods in square confinement, and the formation of a nematic droplet in an isotropic background.
\end{abstract}

\maketitle

\section{Introduction}
Studying phase transitions involves a careful investigation of the free energy of the system, however, computing it can be very hard, even for the simplest interactions \cite{HansenMcDonald}. In the case of symmetry-breaking transitions an order parameter can be defined, which allows one to distinguish a disordered state from an ordered one. From the microscopic Hamiltonian close to a phase transition, a power expansion in terms of this order parameter can often be derived \cite{Altland:2010}, the so-called Landau free energy \cite{Landau:1937}, but it can also be set up phenomenologically based on symmetry grounds. The Landau free energy is usually studied on the level of a saddle-point approximation, such that only polynomial Euler-Lagrange equations have to be solved to understand the phase behaviour, and this procedure has had many successes. Examples include the spontaneous magnetization from a paramagnet to a ferromagnet \cite{Stoof}, the gas-liquid transition \cite{Hohenberg:2015}, and the formation of a superconductor from an ordinary metal \cite{Ginzburg:1950}.


For nematic liquid crystals the order parameter is a traceless and symmetric tensor $\bf Q$ \cite{deGennes:1971} with components $Q_{\alpha\beta}$ where $\alpha,\beta =1,2,3$ in three-dimensional systems. This tensorial form is chosen because the ordered phase breaks rotational ($SO(3)$-) symmetry, but there is a residual $\mathbb{Z}_2$, or up-down symmetry, which requires the theory to be invariant under $O(3)$, rather than $SO(3)$. The bulk Helmholtz (or Gibbs) free energy that describes the first order phase transition towards a nematic phase, can then be expanded as 
\begin{equation}
\Delta F=A(T-T^*)\text{Tr}({\bf Q}^2)-B\text{Tr}({\bf Q}^3)+C(\text{Tr}{\bf Q}^2)^2, \label{eq:thermotropic}
\end{equation}
where $A$, $B$ and $C$ are phenomenological coeffcients, $T$ is temperature and $T^*$ is the temperature of the isotropic spinodal. However, this so-called Landau-de Gennes expansion \cite{deGennes:1971} is only suitable for thermotropics: materials that become liquid crystalline as function of temperature. It has been applied in many situations, ranging from equilibrium \cite{Gramsbergen:1986} to non-equilibrium situations \cite{Yeomans:2002}, including colloidal particles immersed in a thermotropic nematic \cite{Guzman:2003, Nynch:2013} and active nematics \cite{Giomi:2014}. 

In contrast, lyotropic systems, which can consist of hard rods or platelets \cite{Zocher:1925, Bawden:1936, Dogic:1997, Lekkerkerker:1998, Brown:1998}, become ordered as a function of density \cite{Vroege:1992}, and are not described by the free energy of Eq. \eqref{eq:thermotropic}. A simple remedy for this problem would be to replace $T$ in Eq. \eqref{eq:thermotropic} by the density $\rho$, but this cannot capture the density jump that is found in the isotropic-nematic (IN) phase transition, which can be as large as $25\%$  \cite{Lekkerkerker:2000}. Most theories for lyotropics, such as Onsager theory \cite{Onsager:1949}, do exhibit this density jump, but are difficult to handle numerically in more complex situations or geometries, because one has to solve a complicated non-linear integral equation. 

This motivates us to set up a Landau expansion for lyotropics for which we will use the grand potential $\Omega$ rather than the Helmholtz (or Gibbs) free energy $F$ in section II. By using $\Omega$, the expansion parameters will depend on the chemical potential $\mu$ \footnote{Actually, it depends on all intensive variables of the relevant ensemble, in this case not only on $\mu$, but also on $T$. However, we will consider hard particles, hence we do not consider $T$ dependence.}, and the density jump will naturally be encoded through the relation $\partial(\Omega/V)/\partial\mu|_{V,T}=-\rho$, with $V$ the volume of the system and $\rho$ the average density.  {\color{black} Such a Landau expansion in terms of $\bf Q$ is different from, for example, the phase-field-crystal method of Ref. \cite{Wittkowski:2010}, which produces terms that also explicitly depend on density. In our description only a single $\mu$-dependent term is needed to describe the density dependence of the IN transition. It is therefore easier to use than the method proposed in Ref. \cite{Wittkowski:2010}, for which also an Euler-Lagrange equation for $\rho$ needs to be solved, in addition to the one for $\bf Q$.} We will explore the bulk properties of our Landau expansion by fitting and comparing it with Onsager theory \cite{Onsager:1949} in section III, {\color{black} which is exact in the needle limit \cite{Frenkel:1987}}. Afterwards, we fit the square-gradient coefficients by using the hard-rod surface tension on parallel and perpendicular anchoring for a planar IN interface in section IV. This approach is similar to the one that is briefly discussed by Wittmann \emph{et al.} in the context of fundamental measure theory \cite{Wittmann:2014} . However, we will perform a more thorough analysis of the quality of this theory compared to Onsager theory.  Finally, we show some applications: a study of the isotropic core size in a hedgehog defect (section V), linear defects and director textures for rods under confinement (section VI), and the shape and size of a nematic droplet with a homogeneous director field (section VII).

\section{Landau-de Gennes free energy}
\label{sec:ldg}
Let us consider hard rods of length $L$ and diameter $D$ at chemical potential $\mu$ in a macroscopic volume $V$ bounded by a surface $\partial V$. In the Landau grand potential, we will consider terms that depend on $\bf Q$ and its spatial gradient $\nabla \bf Q$,
\begin{align}
&\Delta \Omega[{\bf Q}]=&  \label{eq:LdGfull} \\
& \int_V d{\bf r} \left[\Delta\omega_b({\bf Q}({\bf r}))+\omega_e(\nabla {\bf Q}({\bf r}))\right]+\int_{\partial V} dS \ \omega_s({\bf Q}({\bf r})), \nonumber
\end{align}
where $\Delta\omega_b$ is the bulk grand potential density with respect to the isotropic state, $\omega_e$ describes elastic deformations and surface tension effects, and $\omega_s$ is an anchoring term that describes the interaction with external walls. All the terms should be invariant under ${\bf Q}\rightarrow {\bf U}^T\bf QU$, with ${\bf U}\in O(3)$. We expand the bulk contribution with respect to the isotropic state $\Delta\omega_b$ up until fourth order in ${\bf Q}$, which gives us
\begin{align}
\beta B_2 \Delta\omega_b&({\bf Q}({\bf r});\mu)=\frac{2}{3}a\beta(\mu^*-\mu)Q_{\alpha\beta}Q_{\beta\alpha}- \label{bulk} \\
& \frac{4}{3}b \ Q_{\alpha\beta}Q_{\beta\lambda}Q_{\lambda\alpha}+\frac{4}{9}d \ Q_{\alpha\beta}Q_{\beta\alpha}Q_{\lambda\rho}Q_{\rho\lambda}, \nonumber
\end{align}
where we will use the Einstein summation convention throughout the paper. The second virial coefficient in the disordered isotropic phase is given by $B_2=\pi L^2D/4$ in the limit $L\gg D$, and is included in our definition to render the Landau coefficients $a$, $b$ and $d$ conveniently dimensionless. For simplicity we assume them to be independent of $\mu$. Moreover, $\mu^*$ will turn out to be the chemical potential at the isotropic spinodal. When we assume that the nematic phase is uniaxial, then $\bf Q$ can be expressed in terms of the scalar order parameter $S({\bf r})$ and the director field $\bf n(r)$ for $\alpha,\beta=1,2,3$,
\begin{equation}
Q_{\alpha\beta}({\bf r})=\frac{3}{2}S({\bf r})\left[n_\alpha({\bf r}) n_\beta({\bf r})-\frac{1}{3}\delta_{\alpha\beta}\right].  \label{opa}
\end{equation}
Notice that the largest eigenvalue of ${\bf Q}$ is $S$, while the corresponding (normalized) eigenvector is ${\bf n}$.
Using that $Q_{\alpha\beta}Q_{\beta\alpha}=(3/2)S^2$ and that $Q_{\alpha\beta}Q_{\beta\lambda}Q_{\lambda\alpha}=(3/4)S^3$, we can express Eq. \eqref{bulk} in terms of $S$ as
\begin{equation}
\beta B_2\Delta\omega_b=a\beta(\mu^*-\mu)S^2-bS^3+dS^4. \label{LdGex}
\end{equation}
 
For the terms in gradients of $\bf Q$ we only retain terms up until square gradients in $\bf Q$ which gives us \footnote{To show this, recall that $\partial_\alpha$ transforms as a vector, i.e., $\partial_\alpha\rightarrow U_{\alpha\beta}\partial_\beta$ for ${\bf U}\in O(3)$}
\begin{align}
\beta B_2&\omega_e(\nabla {\bf Q}({\bf r}))=\frac{2}{9}\Big[l_1(\partial_\alpha Q_{\beta\lambda})(\partial_\alpha Q_{\beta\lambda}) \nonumber \\
&+l_2 (\partial_\alpha Q_{\alpha\lambda})(\partial_\beta Q_{\beta\lambda})+l_3(\partial_\alpha Q_{\beta\lambda})(\partial_\lambda Q_{\beta\alpha})\Big], \label{eqelas}
\end{align}
where the dimensionfull parameters $l_1$, $l_2$ and $l_3$ are elastic constants for $\bf Q$. In general, they will depend on $\mu$, but for simplicity, we initially assume them to be constant. Later, in section VII, we will investigate the effect when they are $\mu$-dependent. 

It is instructive to work out $\omega_e$ for the uniaxial case of Eq. \eqref{opa}. Since the norm of the director is a constant and using the vector identities
\begin{equation}
[{\bf n}\times(\nabla\times {\bf n})]_\alpha=-n_\beta\partial_\beta n_\alpha,
\end{equation}
\begin{align}
(\partial_\alpha& n_\beta)(\partial_\alpha n_\beta)=(\nabla\cdot{\bf n})^2+[{\bf n}\cdot(\nabla\times{\bf n})]^2 \\
&+|{\bf n}\times(\nabla\times{\bf n})|^2-\nabla\cdot[{\bf n}(\nabla\cdot{\bf n})+{\bf n}\times(\nabla\times{\bf n})] \nonumber ,
\end{align}
we can recast $\omega_e$ in the form
\begin{widetext}
\begin{align}
&\beta B_2\omega_e=\frac{1}{3}(l_1+l_s/3)|\nabla S|^2+\frac{l_s}{3}({\bf n}\cdot\nabla S)^2+\nabla(S^2)\cdot\left[\left(l_1+\frac{2}{3}l_s\right)(\nabla\cdot{\bf n}){\bf n}+\left(l_1+\frac{1}{3}l_s\right){\bf n}\times(\nabla\times{\bf n})\right]+  \label{eq:elas} \\
&S^2\Big\{(l_1+l_s)(\nabla\cdot{\bf n})^2+l_1[{\bf n}\cdot(\nabla\times{\bf n})]^2+(l_1+l_s)|{\bf n}\times(\nabla\times{\bf n})|^2\Big\} 
-[l_1+(l_s-l_a)/2]\nabla\cdot\Big\{S^2[{\bf n}(\nabla\cdot{\bf n})+{\bf n}\times(\nabla\times{\bf n})]\Big\}, \nonumber 
\end{align}
\end{widetext}
where we introduced $l_s=(l_2+l_3)/2$ and $l_a=(l_2-l_3)/2$. We see that $\omega_e$ encodes for surface tension (first line) and elastic deformations (second line) \cite{Zumer:1998}. Eq. \eqref{eq:elas} can be compared with the Frank elastic free energy $F_e$ \cite{Oseen:1933, Frank:1958} for a bulk nematic phase with a spatially constant bulk order parameter $S_b$,
\begin{align}
&F_e=\frac{1}{2}\int d{\bf r}\{ K_{11} (\nabla\cdot{\bf n})^2+K_{22}({\bf n}\cdot\nabla\times{\bf n})^2 + \label{eq:Frank}
\\ 
&K_{33}|{\bf n}\times(\nabla\times{\bf n})|^2-2K_{24}\nabla\cdot[{\bf n}(\nabla\cdot{\bf n})+{\bf n}\times(\nabla\times{\bf n})]\}, \nonumber
\end{align}
where the terms in the integrand describe splay, twist, bend and saddle splay deformations, respectively. Up to second order in $S_b$ we find $K_{11}=K_{33}=2S_b^2(l_1+l_s)/(\beta B_2)$, $K_{22}=2S_b^2 l_1/(\beta B_2)$ and $K_{24}=S_b^2[l_1+(l_s-l_a)/2]/(\beta B_2)$. Within Onsager theory $K_{33}\gg K_{11}$ and $K_{22}=K_{11}/3$ \cite{Straley:1973}, which is not to be expected to hold in the Landau expansion at this order. Finally, notice that the last term of Eq. \eqref{eq:Frank} is a surface contribution \footnote{Although we say that it is a surface contribution, it does not mean that it cannot have any effect on bulk properties. Actually, it is well known that for certain director field configurations $K_{24}$ can renormalize $K_{11}$, see Ref. \cite{Prinsen:2004}.} and that $l_a$ will typically only contribute to the surface free energy \cite{deGennes:1971}.

Finally, $\omega_s$ in Eq. \eqref{eq:LdGfull} is an anchoring contribution for external walls, for which we assume the Rapini-Papoular (like), or Nobili-Durand form \cite{Rapini:1969,Nobili:1992},
\begin{equation}
\frac{\beta B_2\omega_s}{L}=\frac{w}{2}\left[Q_{\alpha\beta}({\bf r})-Q_{\alpha\beta}^0({\bf r})\right]^2,
\end{equation}
with $w$ the dimensionless anchoring strength and ${\bf Q}^0$ the preferred value of $\bf Q$ on the surface.
Although it is possible to minimize $\Delta\Omega$ with respect to $S$, it is more convenient and more general to directly minimize Eq. \eqref{eq:LdGfull} with respect to $\bf Q$ when the director field varies as function of position. However, one has to perform the minimization under the constraint that $\bf Q$ is traceless and symmetric, see for example Ref. \cite{Sluckin:1987}.
\section{Bulk properties}
\label{sec:bulk}
\begin{figure}[ht]
\includegraphics[width=0.45\textwidth]{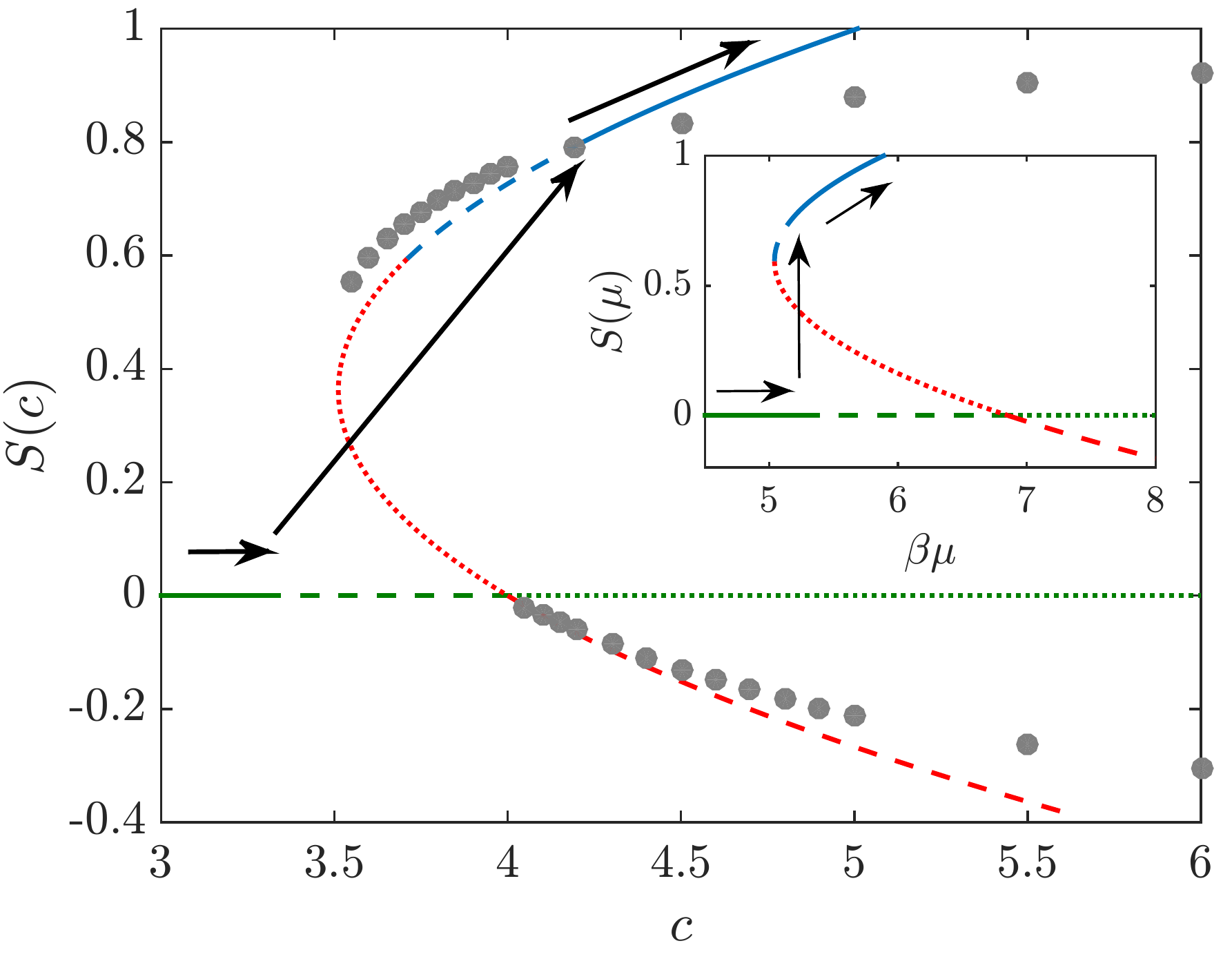}
\caption{Bifurcation diagram for the lyotropic Landau-de Gennes (LdG) theory in the density-order parameter $(c,S)$ representation and in the chemical potential-order parameter $(\mu,S)$ representation (inset) by fitting it to coexistence data from Onsager theory. We use green for the isotropic branch $S_I$, blue for the upper nematic branch $S_N^+$, and red for the lower nematic branch $S_N^-$. We use full lines whenever the respective branch is (globally) stable, dashed lines whenever they are metastable and dotted lines whenever they are absolutely unstable. In grey we show data points obtained from Onsager theory and we see that the LdG theory matches well with it. With arrows we indicate the isotropic-nematic transition in both representations, exhibiting a density jump in the $(c,S)$ representation when the ordered phase starts to form.}
\label{fig:bifurcation}
\end{figure}

In bulk, we assume a fully uniaxial nematic phase and hence it is sufficient to investigate Eq. \eqref{LdGex}. The (meta)stable and unstable points are found by the condition $\partial\Delta\omega_b/\partial S=0$, resulting in the following solutions
\begin{align}
S_I&=0, \\
S_N^\pm&=\frac{3b}{8d}\left(1\pm\sqrt{1-\frac{32 ad \beta(\mu^*-\mu)}{9b^2}}\right).
\end{align}
The stability of these points can be investigated by analyzing the sign of $\partial^2\Delta\omega_b/\partial S^2$. The isotropic spinodal $\mu^*$ is defined by $\partial^2\Delta\omega_b/\partial S^2|_{S=S_I}=0$, while the nematic spinodal $\beta\mu^+=\beta\mu^*-9b^2/(32ad)$ is the $\mu$ for which $\partial^2\Delta\omega_b/\partial S^2|_{S=S_N^+}=0$. Finally, the binodal $\beta\mu_{IN}=\beta\mu^*-b^2/(4ad)$ is determined from $\Delta\omega_B(S_I)=\Delta\omega_B(S_N^+)$. A stability analysis shows that (i) for $\mu<\mu^+$ the isotropic phase is the stable configuration, (ii) for $\mu^{+}<\mu<\mu_{IN}$ we have that $S_N^-$ is absolutely unstable, while $S_N^+$ is metastable and the isotropic phase is stable, and  (iii) for $\mu_{IN}<\mu<\mu^{*}$ we have that $S_N^-$ is absolutely unstable, while $S_N^+$ is stable and the isotropic phase is metastable. (iv) For $\mu>\mu^{*}$ we have that $S_N^-$ is metastable, while $S_N^+$ is stable and the isotropic phase is unstable. The resulting bifurcation diagram listing the stability of all these branches is shown in the inset of Fig. \ref{fig:bifurcation}, where we indicate with arrows the IN transition as function of $\mu$. Below, we derive the values of the coefficients we used in Fig. \ref{fig:bifurcation} from fits to Onsager theory.

When $S$ is known for a given $\mu$, one can convert $\mu$ to the dimensionless density $c=B_2\rho$. For this we introduce the grand potential density of the isotropic state $\omega_I$ and define $\omega:=\omega_I+\Delta\omega_b$. Then we find $\partial(B_2\omega)/\partial\mu=-c$, such that
\begin{equation}
c(\mu)=c_I(\mu)+a S^2, \label{eq:densS}
\end{equation}
where we have defined $c_I(\mu)=-\partial(B_2\omega_I)/\partial\mu$. Within Onsager theory, we calculate $\omega_I$ by using an isotropic distribution function, such that $\beta\mu(c_I)=\log(c_I/4\pi)+2c_I$ \cite{Roij:2005}. By inverting this relation, one obtains $c_I(\mu)$. Together with Eq. \eqref{eq:densS}, and $S$ one can determine $c$.

At isotropic-nematic coexistence, we have from Eq. \eqref{eq:densS} and the analysis above that
\begin{align}
c(\mu_{IN})=c_I(\mu_{IN})+a S_{IN}^2,  \\
\beta\mu_{IN}=\beta\mu^*-\frac{b^2}{4ad}, \\
S_{IN}=\frac{b}{2d}. 
\end{align}
Within Onsager theory, it is known that \cite{Roij:2005} $c_I(\mu_{IN})=3.290$, $c(\mu_{IN})=4.191$, $\beta\mu^*=6.855$, $\beta\mu_{IN}=5.241$ and $S_{IN}=0.7922$. Using these values, we find $a=1.436$, $b=5.851$ and $d=3.693$. With this set of parameters, we determine that $\beta\mu^+=5.039$ \footnote{Notice that we could also have used $\mu^+$ to determine the Landau coefficients. However this quantity is hard to determine numerically from Onsager theory.}.  We plot the bifurcation diagram in Fig. \ref{fig:bifurcation}, indicating the stable, metastable and unstable regions in the $(c,S)$ representation and in the inset we show the $(\mu,S)$ representation. Arrows indicate the IN transition in both representations, for which the density jump is correctly captured by construction. We also give a comparison with Onsager theory (the grey circles) for which the bifurcation diagram is known \cite{Kayser:1978, Roij:2005}.

We note that $S_N^+>1$ for $c\gtrsim5$ in Fig. \ref{fig:bifurcation}, which is unphysical. A simple remedy for this problem would be to replace $S^4\rightarrow S^4/(1-S)$ in Eq. \eqref{LdGex}, which ensures that $S\leq1$. However, this complicates the free energy, especially when the full {\bf Q}-tensor theory is needed, so it will not be considered here. We have seen in our calculations that this remedy does give better results for the nematic branches at high densities when compared with Onsager theory. However, our calculations show that inhomogeneous Landau theories are less accurate for these expansions, presumably because accuracy in the metastable regime is more important than in the high-density regime. 

\section{Isotropic-Nematic interface}
\label{sec:flat}
\begin{figure}[t]
\includegraphics[width=0.5\textwidth]{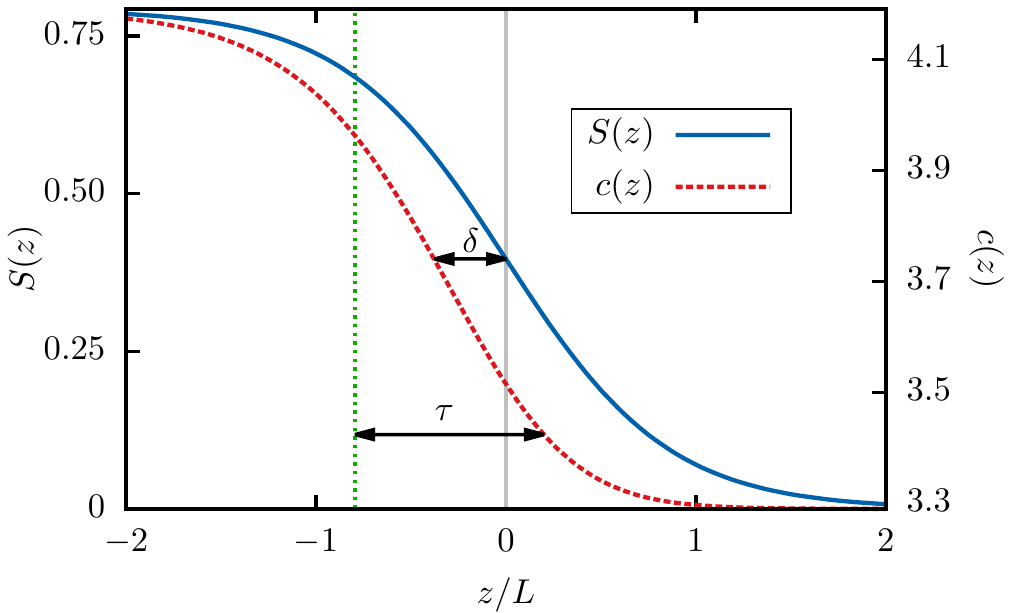}
\caption{Interfacial profiles $S(z)$ and $c(z)$ between an isotropic phase and a nematic phase, calculated within the lyotropic LdG theory. Two characteristic distances are defined to quantify the shape of the profiles: $\delta$ is the displacement between the center of the $S$ profile and that of the $c$ profile, where we defined a measure for the interfacial width $\tau=|z_+-z_-|$, with $c'''(z_\pm)=0$. }
\label{fig:interface}
\end{figure}
Let us now consider an inhomogeneous system that for $z\rightarrow\infty$ consists of an isotropic fluid, while for $z\rightarrow-\infty$ there is a bulk nematic with order parameter $S_b$. {\color{black} For simplicity, we neglect biaxial effects within our LdG theory, which can be important, but can be included quite easily if necessary \cite{Wheeler:1997, Kamil:2009}.} Moreover, we assume homogeneity in the plane perpendicular to the $z$ axis. As a natural consequence a planar interface will develop between the two bulk phases, with an order parameter profile $S(z)$ and a density profile $c(z)$, that can be calculated within our Landau theory. The surface tension for parallel $\gamma_\parallel$ and perpendicular anchoring $\gamma_\perp$ of such a system are known within Onsager theory \cite{Shundyak:2006} and these quantities will be used to fit the constants $l_1$ and $l_s=(l_2+l_3)/2$ from Eq. \eqref{eqelas}. We fix the director field $\bf n$ to be spatially constant with a specified orientation and let $\alpha$ be the angle of the director with the interface normal. This means that for $\alpha=0$ ($\alpha=\pi/2$) the rods are aligned perpendicular (parallel) to the interface. With this definition, we can write ${\bf n}\cdot\nabla S = S'(z)\cos\alpha $, where henceforth a prime denotes differentiation with respect to $z$.

The Landau grand potential per unit interfacial area $A$ for this geometry is
\begin{equation}
\frac{\Delta\Omega[S]}{A}=\int dz\left[\frac{m(\alpha)}{2\beta B_2} \left(S'(z)\right)^2+\Delta\omega_b(S(z);\mu)\right], \label{eq:grandpotint}
\end{equation}
with stiffness constant $m(\alpha)=(2/3)[l_s\cos^2\alpha+l_1+l_s/3]$. 
It is straightforward to rewrite the Euler-Lagrange equation $\delta\Delta\Omega_\text{LdG}[S]/\delta S=0$, into
\begin{align}
m(\alpha)S''(z)=\frac{\partial[\beta B_2\Delta\omega_b(S(z);\mu)]}{\partial S(z)}, \nonumber \\
\lim_{z\rightarrow-\infty}S(z)=S_b, \  \lim_{z\rightarrow\infty}S(z)=0.
\end{align}
For our purposes it suffices to consider this equation at coexistence, where $\mu=\mu_{IN}$ and $S_b=S_{IN}$. Multiplying this equation with $S'(z)$ and integration over $z$ gives
\begin{equation}
\frac{m(\alpha)}{2}(S'(z))^2=dS^2(S-S_{IN})^2, \label{eq:melle}
\end{equation}
where we used that $\beta B_2\Delta\omega_b(S(z);\mu_{IN})=dS^2(S-S_{IN})^2$, and where an integration constant vanishes because $S'(z)\rightarrow 0$ for $z\rightarrow\pm\infty$. Taking the square root of this equation and choosing the positive root since it is consistent with our boundary conditions, it is straightforward to find that the order parameter profile reads
\begin{equation}
S(z)=\frac{S_{IN}}{2}\left[1-\tanh\left(\frac{z}{2\xi}\right)\right],
\end{equation}
after introducing the correlation length $\xi=[2dm(\alpha)]^{1/2}/b$. The profile for $c(z)$ can be obtained from Eq. \eqref{eq:densS}, as
\begin{equation}
c(z)=c_I+\frac{a S_{IN}^2}{4}\left[1-\tanh\left(\frac{z}{2\xi}\right)\right]^2.
\end{equation}

By definition $\Delta\omega_b(S_{IN},\mu_{IN})=0$, which means that the bulk pressure comes entirely from the isotropic contribution to the grand potential, $\omega_I=-\beta B_2p_{IN}$. Hence, $\Delta\Omega/A$ is the surface tension when evaluated at coexistence,
\begin{align}
\beta B_2 \gamma(\alpha)=m(\alpha)\int_{-\infty}^\infty dz [S'(z)]^2=S_{IN}^3\sqrt{\frac{dm(\alpha)}{18}}  \label{eq:surftens},
\end{align}
where we used Eq. \eqref{eq:melle}. Using Eq. \eqref{eq:surftens} we find that $m(\alpha)=9/(8d)[\pi\beta\gamma(\alpha) LD/S_{IN}^3]^2L^2$, and the surface tensions known from Onsager theory, $\beta\gamma_\parallel LD=0.156$ ($\alpha=\pi/2$) and $\beta\gamma_\perp LD=0.265$ ($\alpha=0$) \cite{Shundyak:2006}, are used as input parameters. We thus find $m_\parallel=0.296 L^2$, and $m_\perp=0.854 L^2$. For the perpendicular case ($\alpha=0$), we plot the profiles $c(z)$ and $S(z)$ in Fig. \ref{fig:interface}, showing that the density profile is shifted with respect to the order parameter profile, which is consistent with Onsager theory and simulations \cite{Shundyak:2006}. The same phenomenon is found for $\alpha=0$, but the interfaces have a smaller width. {\color{black} This shift seems to be robust, general and universal as several types of theories yield similar predictions for its sign and magnitude, even for the isotropic-plastic crystal interface \cite{Praetorius:2013}.} 

For the above obtained values of $m_\parallel$ and $m_\perp$, it follows using the relation below Eq. \eqref{eq:grandpotint} that $l_1=0.165L^2$ and $l_s=0.837L^2$. In this case it is not possible to determine $l_a$. Using these values, we find $K_{11}/K_{22}=6$, which should be contrasted with the exact relation where this ratio should be equal to 3 for $L\rightarrow\infty$ \cite{Straley:1973, Poniewierski:1979, Stecki:1980}. 

To assess the quality of our calculations, we introduce two characteristic lengths. The first one is defined as $\delta=|z_S-z_c|$, where $S(z_S)=S_{IN}/2$ and $c(z_c)=[c(\mu_{IN})+c_I(\mu_{IN})]/2$, hence we see that $\delta$ is a measure for the shift of $c(z)$ with respect to $S(z)$. Another length scale is the width of $c(z)$, which can be defined as $\tau=|z_+-z_-|$, where $z_\pm$ satisfies $c'''(z_\pm)=0$. For $\alpha = 0$, we find that $\delta_\parallel=0.223L$ and $\tau_\parallel=0.586L$, while within Onsager theory the values are $\delta_\parallel^O=0.45L$ \cite{Shundyak:2006} and $\tau_\parallel^O=0.697L$ \cite{Shundyak:2003}. Finally, for $\alpha=\pi/2$, we have determined that $\delta_\perp=0.378L$ and $\tau_\perp=0.994L$, however, the values of $\delta_\perp^O$ and $\tau_\perp^O$ are not reported in the literature. Where comparisons are possible, we do see that the LdG results compare quite favourably with Onsager theory, where one should keep in mind the enormously reduced numerical effort of the LdG theory compared to the Onsager theory of the IN interface. 

\section{Radial hedgehog defect}
\label{sec:hedg}
\begin{figure}[t]
\includegraphics[width=0.5\textwidth]{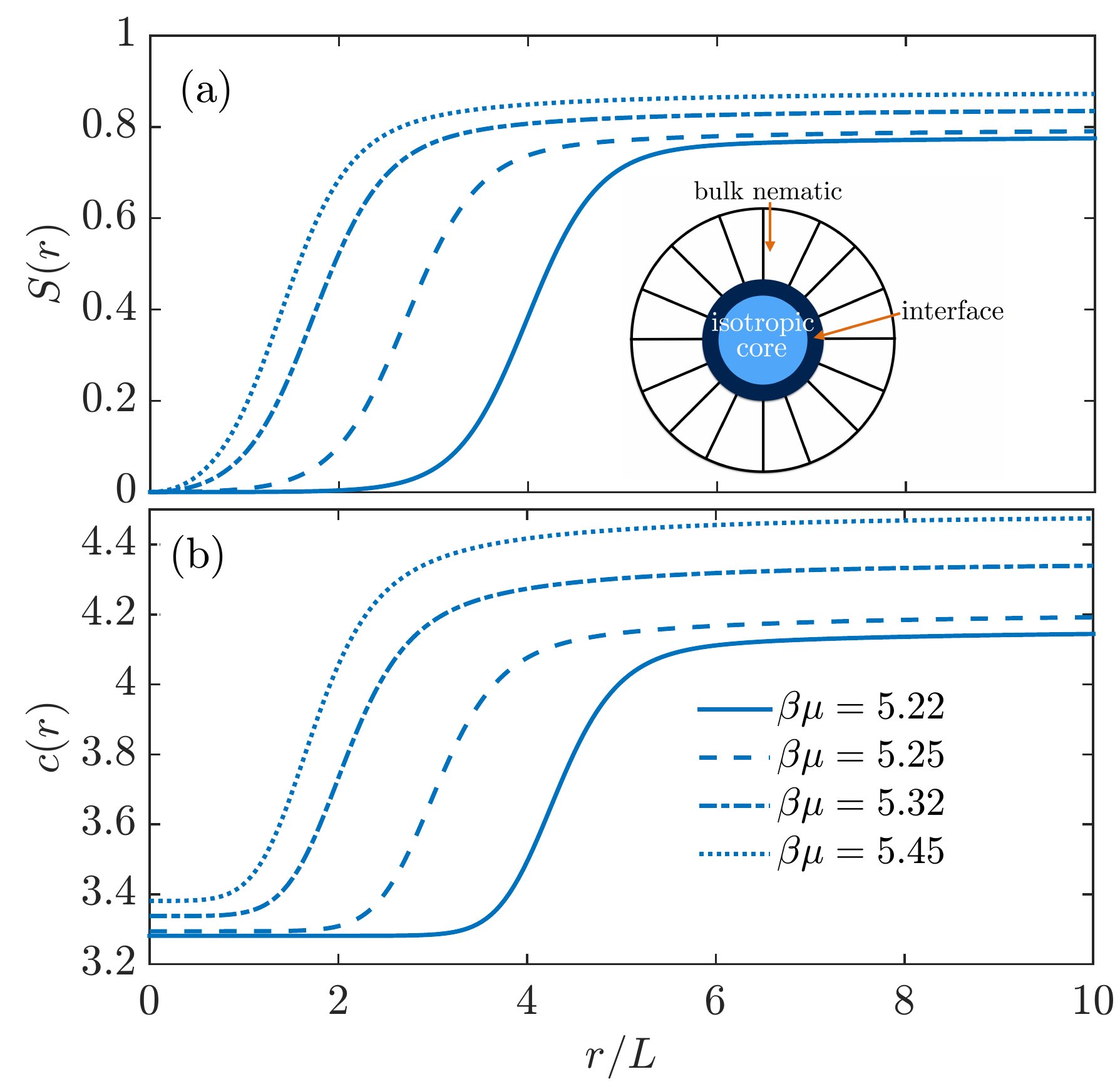}
\caption{Internal structure of a hedgehog defect for various chemical potentials $\mu$. In (a) the order parameter profile $S(r)$ is shown and in the inset we show a schematic top view of the defect. In (b) we plot the dimensionless local density $c(r)$. The profiles clearly show the structure of an isotropic core centered at $r=0$.}
\label{fig:hedgehog}
\end{figure}

In this section, we will study the hedgehog defect, an object that has received much attention in the thermotropic liquid crystal literature \cite{Schopohl:1988, Greco:1992, Mkaddem:2000, Majumdar:2012}, but for which little is known for lyotropic liquid crystals. To study this type of defect, we assume locally uniaxial symmetry and consider
\begin{equation}
{\bf Q} ({\bf r})= \frac{3}{2}S({\bf r})\left({\bf e}_r\otimes{\bf e}_r^T-\frac{1}{3}\mathbb{I}\right),
\end{equation}
where ${\bf n(r)}={\bf e}_r$ is the radial unit-vector. If the hedgehog defect would consist of a bulk nematic phase, the elastic free energy density diverges at the centre $r=0$, and the only way for the system to lower its free energy is by a melting transition of the core to an isotropic phase. Landau theory allows us to determine the internal structure of the resulting defect core, shown schematically in the inset of Fig. \ref{fig:hedgehog}. This calculation would not be possible within continuum theories such as e.g. Eq. \eqref{eq:Frank}, in which the spatial variation of $S({\bf r})$ is ignored. In such Frank-Oseen theories, a cut-off length is needed to assess the size of the isotropic core, whereas the core size will follow naturally from the regime where $S({\bf r})$ vanishes within our LdG theory. Finally, the rod length $L$ is a natural length scale in our calculations, because we determined the stiffness constants $m_\parallel$ and $m_\perp$ from a microscopic theory. This allows us to estimate the isotropic core size in terms of $L$.

\begin{figure}[t]
\includegraphics[width=0.5\textwidth]{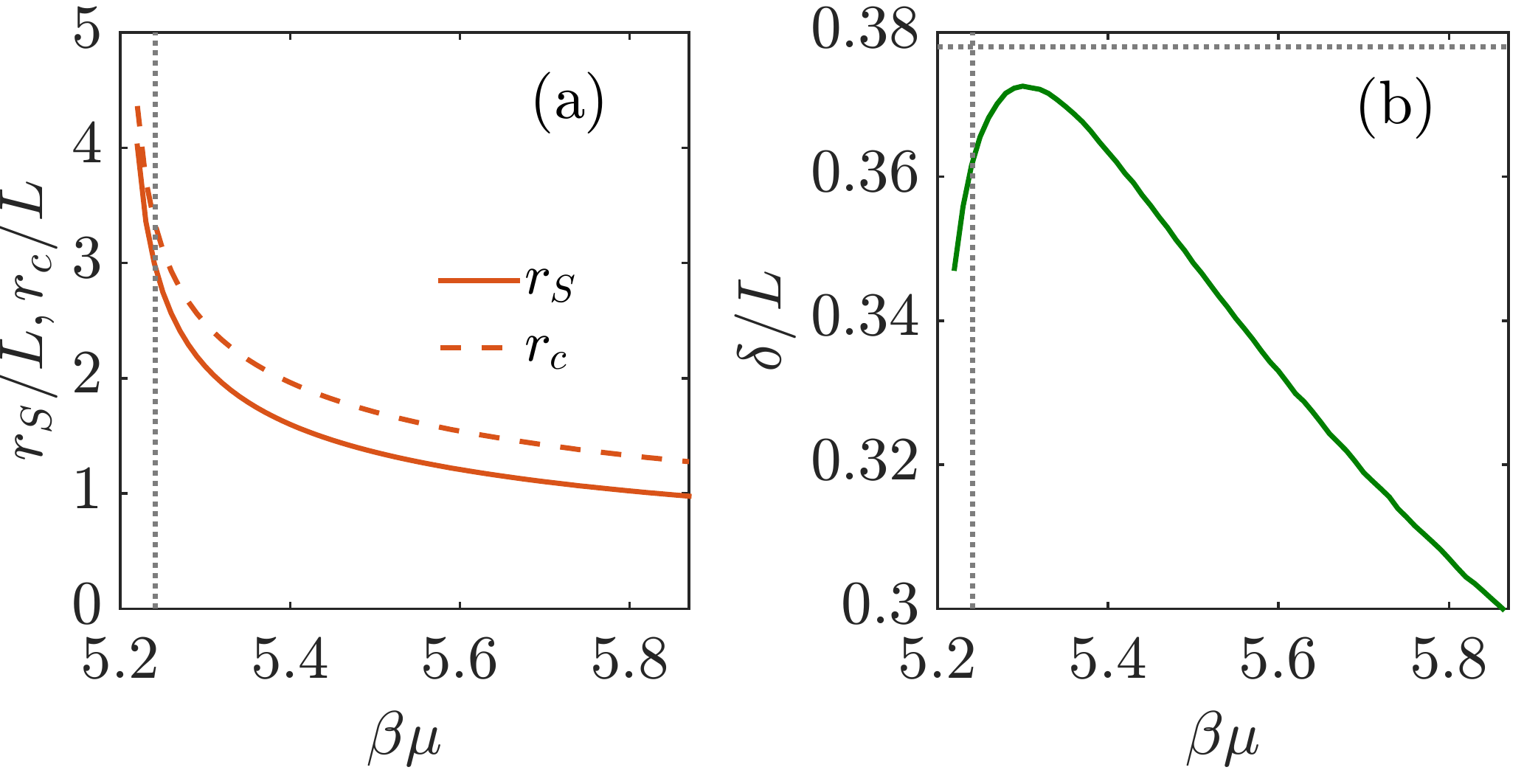}
\caption{(a) Isotropic core size measured by the positions $r_S$ and $r_c$ for which the order parameter $S$ and the density $c$ reach the average of their minimum and maximum value, respectively. (b) The interfacial shift $\delta=|r_S-r_c|$ as function of $\mu$. The vertical dotted lines indicate the bulk binodal $\mu=\mu_{IN}$, and the horizontal line in (b) is the shift $\delta_\perp$ in a planar geometry with a fixed director field perpendicular to the interfacial plane, as determined in section IV.}
\label{fig:isosizedelta}
\end{figure} 

Since we fix the director field to be radial, it suffices to evaluate Eq. \eqref{LdGex} and \eqref{eq:elas}. A radial director field is irrotational, $\nabla\times{\bf e}_r={\bf 0}$, and has a non-vanishing divergence, $\nabla\cdot{\bf e}_r=2/r$. From Eq. \eqref{eq:LdGfull} and Eq. \eqref{eqelas} it is then straightforward to derive the grand potential 
\begin{align}
\beta B_2&\Delta\Omega[S]=\int d{\bf r}\Bigg[\frac{1}{3}\left(l_1+\frac{4l_s}{3}\right)(\partial_r S)^2  \\
&+\partial_r(S^2)\frac{2}{r}\left(l_1+\frac{2}{3}l_s\right)+S^2(l_1+l_s)\frac{4}{r^2}+\beta B_2\Delta\omega_b\Bigg] \nonumber,
\end{align}
for which the Euler-Lagrange equation is
\begin{equation}
S''(r)+\frac{2}{r}S'(r)-\frac{6}{r^2}S(r)=\frac{3\beta B_2}{2 m_\perp}\frac{\partial\Delta\omega_b}{\partial S}, \label{eq:hedgehogEL}
\end{equation}
to be solved for the boundary conditions $S(0)=0$ and $S(R)=S_N^+(\mu)$. The radius $R$ is the size of the hedgehog defect, which can be thought of as the radius of a finite spherulite or a bulk nematic where there is a finite region of size $R$ where the rods are radially aligned. In our calculations we set $R=50L$. Notice that the surface term in Eq. \eqref{eqelas} does not contribute because we consider strong homeotropic anchoring conditions at $r=R$.

Solving Eq. \eqref{eq:hedgehogEL} for a given $\mu$ gives the structure of a hedgehog defect $S(r)$, shown in Fig. \ref{fig:hedgehog}(a) for several $\mu$. Indeed, an isotropic core centered at the origin is found. For $r\downarrow 0$, the profiles behave locally as $S(r)=\mathcal{O}(r^2)$ and for $r\rightarrow\infty$ as $S(r)=S_N^+(\mu)-\mathcal{O}(1/r^2)$, as is well known \cite{Schopohl:1988}. Moreover, increasing $\mu$,  which is equivalent to setting a higher bulk density, gives rise to a smaller isotropic core size. This behaviour is also observed in thermotropic Landau theory \cite{Schopohl:1988}, however there are two new features that our version provides. Firstly, as was mentioned earlier, $m_\perp$ is determined by a fit to the Onsager result, and hence the rod length $L$ is an intrinsic (microscopic) length scale of the theory. This allows us to determine the isotropic core size in terms of $L$. We see that the isotropic core size is always $\mathcal{O}(L)$ sufficiently far from the nematic spinodal, in accordance with experiments \cite{Zasadzinski:1986}. This is in sharp contrast with thermotropic liquid crystals, where the isotropic core size is macroscopic in size compared to the microscopic size of the molecules, which have a length on the order of $10^{-9}$ m \cite{Schopohl:1988}. The reason is that for thermotropics, the binodal lies always close to the isotropic spinodal temperature $T^*$ and the nematic spinodal temperature $T_+$, since typically $T_+-T^*\lesssim1$ K \cite{Gramsbergen:1986}. Secondly, from $S(r)$ we can determine the density profiles $c(r)$, for which we show some examples in Fig. \ref{fig:hedgehog}(b) for the same set of $\mu$. Similar to the planar interface case, $c(r)$ is shifted with respect to $S(r)$ towards the region where the bulk nematic phase is found.

The isotropic core size can be characterized by $r_S$ or $r_c$, the positions for which $S(r)$ and $c(r)$ attains the average of their minimum and maximum value, respectively, which we show as a function of $\mu$ in Fig. \ref{fig:isosizedelta}(a). We see that both quantities increase with decreasing $\mu$, showing that at low (bulk) densities the rods have a lower tendency to order, which facilitates an isotropic phase. When $\mu$ approaches the nematic spinodal $\mu_+$, the core size diverges, because the nematic phase becomes absolutely unstable. However, it is hard to determine numerically the exact state point for which this happens, because of convergence problems in Eq. \eqref{eq:hedgehogEL} for $\beta\mu\lesssim 5.22$. 

We plot the interfacial shift $\delta=|r_S-r_c|$ in Fig. \ref{fig:isosizedelta}(b), revealing a weak (but non-monotonic) variation with $\mu$. As a comparison we indicate $\delta_\perp$, which is the result found in the flat geometry of Sec. \ref{sec:flat}, as the dotted horizontal line. We do not make a comparison with the interfacial width $\tau$ since it is an ill-defined quantity for the hedgehog defect: the equation $c'''(r)=0$ has only one solution. Finally, we remark that the hedgehog defect can be unstable towards a ring disclination or a split core defect, as was found for thermotropic LdG theory \cite{Mkaddem:2000}. For hard rods the split core defect has been realized by applying an external magnetic field \cite{Otten:2012}. Ring disclinations, however, are expected not to occur in hard-rod systems, because, in contrast to molecular systems, we have $K_{11}\gg K_{33}$.

\section{Confined hard rods}
\label{sec:rods}
An application for which we need the full $\bf Q$-tensor theory is the confinement of rods in a rectangular cuboid. Such a system has been investigated using a mean-field Onsager model and Frank-Oseen model \cite{Lewis:2014}, but also using Monte-Carlo simulations \cite{Garlea:2015} and within Landau theory \cite{Tsakonas:2007}. In the latter case, however, local density variations were \emph{not} considered.

\begin{figure}[t]
\includegraphics[width=0.5\textwidth]{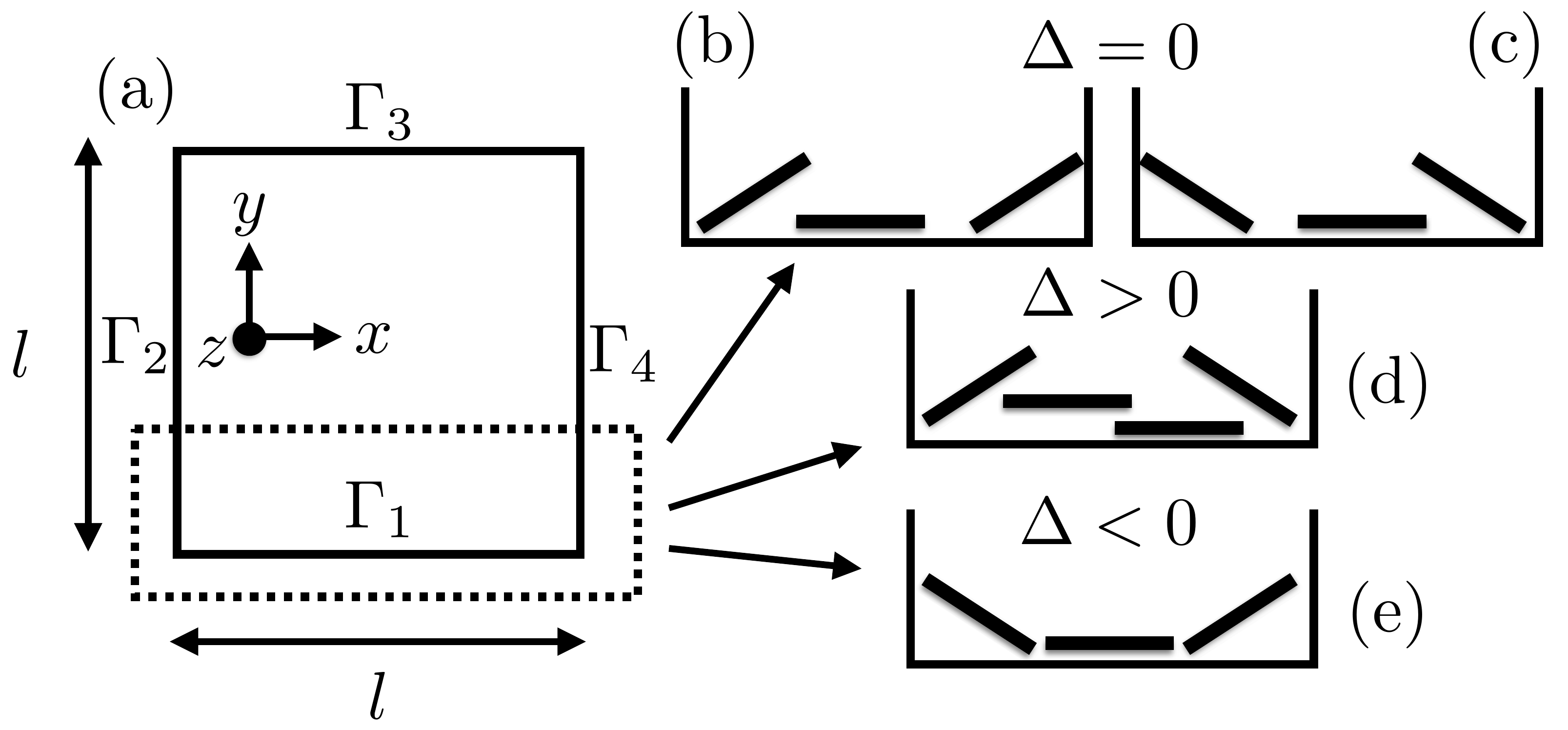}
\caption{(a) A rectangular cuboid with a square base of dimensions $l\times l$ in which we confine rods with diameter $D$ and length $L$. Various anchoring boundary conditions at the four walls $\Gamma_1,...,\Gamma_4$ are investigated. (b)-(c) For $\Delta=0$ (see Eq. \eqref{eq:directorboundary}) the rods at the corners of the square have the same angle with the bottom wall, while (d) for $\Delta>0$ the rods always move to the edge farthest from the corner towards the opposing wall or (e) for $\Delta<0$ the edge closest to the corner is moved towards the opposing wall. }
\label{fig:sqgeometry}
\end{figure}

Let us consider a rectangular cuboid with a square base of dimensions $l\times l$ in the $xy$-plane as illustrated in Fig. \ref{fig:sqgeometry}(a).  To simplify the problem, we take the height equal to the diameter of the rods, such that the rods will necessarily order within the $xy$-plane. We implement this by setting $Q_{zz}=-1/2$ and $Q_{zx}=Q_{zy}=0$. Following Ref. \cite{Garlea:2015}, we use the largest positive eigenvalue $\lambda_+$ of $\bf Q$ to determine the degree of order, while the corresponding eigenvector ${\bf n}$ is a measure for the alignment in the $xy$-plane. Observe that $\lambda_+$ is not always the same as the scalar order parameter $S$, which is defined as the absolute largest eigenvalue.
\begin{figure}[t]
\includegraphics[width=0.51\textwidth]{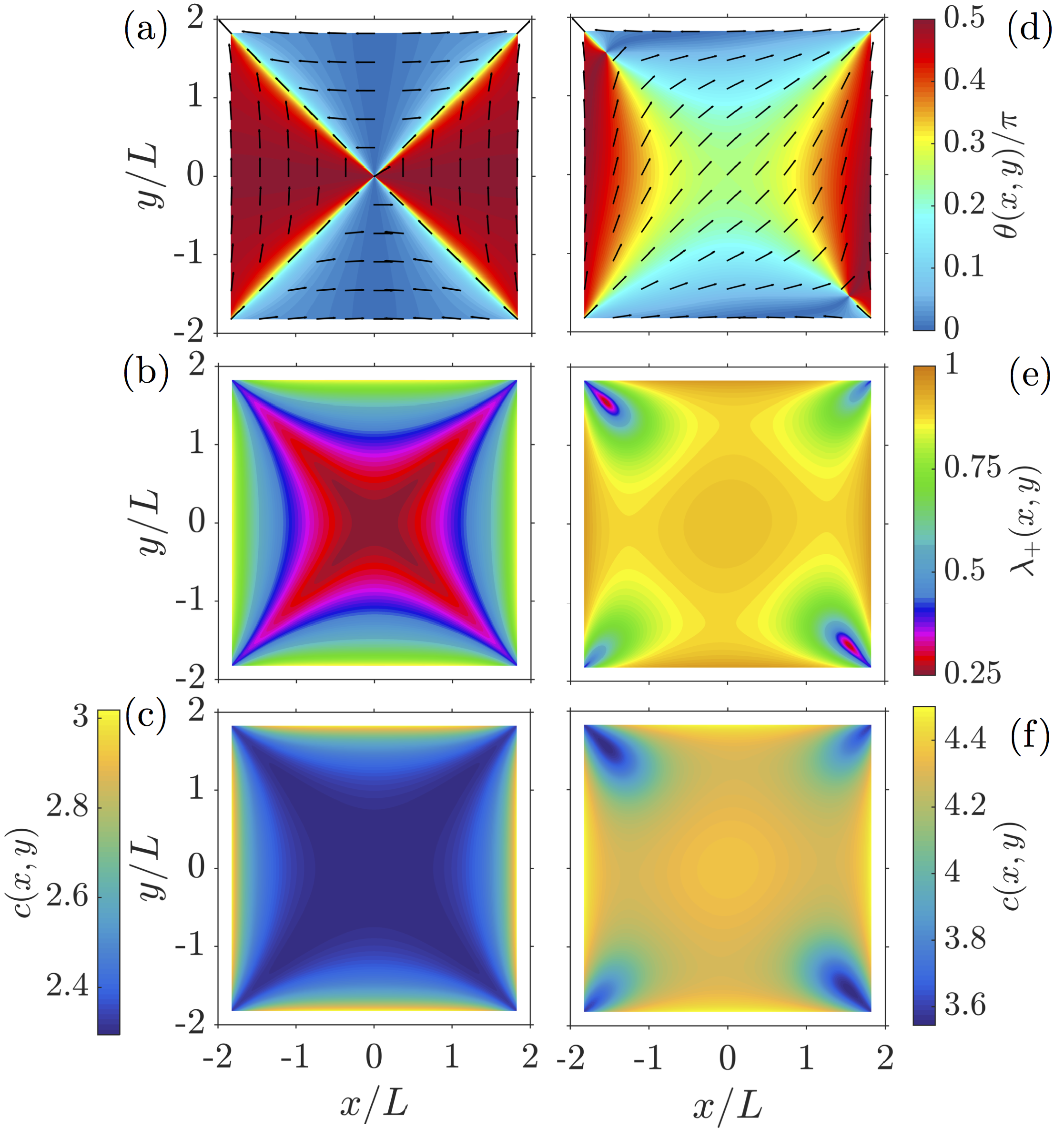}
\caption{Profiles of the director angle $\theta(x,y)$ (top), eigenvalue $\lambda_+(x,y)$ (middle) and density $c(x,y)$ (bottom) in a cell of $73/20\times 73/20$ in units of rod length $L$ and where the height of the slab is chosen to be equal to the diameter $D$ of the rods. We fix the anchoring strength $w=10$ and anchoring parameter $\Delta=0.1$, and investigate their effects on an isotropic state (a)-(c) at chemical potential $\beta\mu=2$, while in (d)-(f) we consider a nematic state $\beta\mu=5$. The bulk 2D IN transition occurs at $\beta\mu_{IN}=2.03$ as explained in the text. The ordering effects of the walls can be seen in (a) where $\theta$ is the angle with the $x$ axis, while the degree of ordering is shown in (b) and (e), and the density $c$ in (c) and (f).}
\label{fig:defects1}
\end{figure}
We parametrize
\begin{equation}
{\bf Q}(x,y)=\begin{pmatrix} 
\frac{1}{4}+q_1(x,y) & q_2(x,y) & 0 \\
q_2(x,y) & \frac{1}{4}-q_1(x,y) & 0 \\
0 & 0 & -\frac{1}{2} 
\end{pmatrix},
\end{equation} 
for which we can derive (see Appendix A) the Euler-Lagrange equations for ${\bf q}=(q_1,q_2)$, 
\begin{equation}
(l_1+l_s)\nabla^2{\bf q}=3a\beta(\mu^*-\mu){\bf q}-\frac{9}{2}b{\bf q}+\frac{d}{2}(3+16q_1^2+16q_2^2){\bf q}, \label{eq:EL1}
\end{equation}
with the boundary conditions
\begin{equation}
(l_1+l_s)(\hat{\boldsymbol{\nu}}\cdot\nabla {\bf q})+\frac{9}{4}w({\bf q}-{\bf q}^0)=0, \label{eq:EL2}
\end{equation}
and $\hat{\boldsymbol{\nu}}$ an outward pointing normal vector. Unless stated otherwise, we assume relatively strong anchoring conditions, $w=10$. Moreover, we assume anchoring conditions at the four walls of the form
\begin{equation}
{\bf Q}^0({\bf r})=\frac{3}{2}\left[{\bf n}^0({\bf r})\otimes\left({\bf n}^0\right)^T({\bf r})-\frac{1}{3}\mathbb{I}\right], \quad {\bf r}\in\Gamma_i,
\end{equation} 
with a specified director ${\bf n}^0$ and $\Gamma_i$ $(i=1,..,4)$ indicating the four walls. In general, we will assume planar anchoring at the four walls, but we also want to investigate various director configurations at the four corners. For example, for $(x,y)\in\Gamma_1$ we assume
\begin{align}
n_x^0&=\sqrt{1-\Delta^2\sin^2\left(\frac{\pi x}{l}\right)}, \nonumber \\
n_y^0&=-\Delta\sin\left(\frac{\pi x}{l}\right), \label{eq:directorboundary}
\end{align}
with anchoring paramater $\Delta$, which allows us to study metastable states that are otherwise hard to access. For $\Delta=0$ this gives an equal weight to either of the two configurations shown in Fig. \ref{fig:sqgeometry}(b) and (c), whereas a bias in the configurations of the director occurs when $\Delta\neq0$. For $\Delta>0$ the configuration of Fig. \ref{fig:sqgeometry}(d) is preferred and for $\Delta<0$ the configuration of Fig. \ref{fig:sqgeometry}(e). Analogous expressions can be derived for the other three boundaries which we can summarize for $(x,y)\in\Gamma_i$ $(i=1,2,3,4)$ as
\begin{align}
q_1^0&=(-1)^i\frac{3}{4}\left[2\Delta^2\sin^2\left(\frac{\pi x_i}{l}\right)-1\right], \label{eq:boundq1} \\
q_2^0&=-\frac{3}{2}\Delta\sin\left(\frac{\pi x_i}{l}\right)\sqrt{1-\sin^2\left(\frac{\pi x_i}{l}\right)} ,\label{eq:boundq2}
\end{align}
with $(x_1,x_2,x_3,x_4)=(x,y,-x,-y)$.
Solving the set of equations Eq. \eqref{eq:EL1}, \eqref{eq:EL2}, \eqref{eq:boundq1} and \eqref{eq:boundq2} gives $\lambda_+(x,y)$ and ${\bf n}(x,y)$ \footnote{Notice that {\bf n} is strictly speaking only a well-defined quantity whenever $\lambda_+>1/4$}. Moreover, within our Landau theory we can extract the density $c(x,y)=c_I(\mu)+(2a/3) \text{Tr}({\bf Q}^2)$. Interestingly, it turns out that $c(x,y)$ is a good measure for the local two-dimensional density $\rho_{2D}=N/A$, with $N=\rho A D$ the number of rod and $A=l^2$ the base area, such that $c=(\pi/4)L^2\rho_{2D}$, is a natural dimensionless real density.

 {\color{black} Note that the present square-gradient theory cannot account for layering at a hard wall, which in principle is present in simulations and theory of these systems \cite{Mao:1997, Dijkstra:2001} although not in a pronounced way given the large particle aspect ratios considered. The extension of the Landau theory to include, for example, a smectic phase in the case of short rods to describe pre-smectic layering for perpendicular anchoring is, however, straightforward \cite{McMillan:1972}.}

In Fig. \ref{fig:defects1} we show examples of the structures that can be found for $\Delta=0.1$ and $l/L=73/20$. The latter system size is chosen because it enables us to compare our results with the simulations of Ref. \cite{Garlea:2015}. We consider $\beta\mu=2$ (2D isotropic) and $\beta\mu=5$ (nematic), which should be compared with $\beta\mu_{IN}^{2D}=\beta\mu^*-(3b-d)/(2a)=2.03$ where the bulk phase transition in this 2D geometry takes place according to our 3D-based LdG theory. In the isotropic state for $\Delta=0.1$, the rods align along the four walls, while pointing their endpoints towards the corners along the diagonals, as can be seen in Fig. \ref{fig:defects1}(a) where we plot the (minimal) angle $\theta$ of ${\bf n}$ with respect to the the $x$-axis. Fig. \ref{fig:defects1}(b) shows that the core is 2D isotropic since $\lambda_+=1/4$, but close to the walls ordering is induced, $\lambda_+>1/4$. Furthermore, the density is lowest where the ordering is the smallest, see Fig. \ref{fig:defects1}(c). However, a complete study of the wetting and pre-wetting properties of the walls is out of the scope of this paper, since a more sophisticated surface free energy may be needed \cite{Sullivan:1987}. In the nematic state, there is a competition between the ordering effects of the wall and the density-induced order due to the rod-rod interactions. This results in a lens-shaped director field structure (Fig. \ref{fig:defects1}(d)), and for this specific boundary condition it results in defects near two of the four corners with an isotropic core shown in red in Fig. \ref{fig:defects1}(e), with a reduction in local density shown in blue in Fig. \ref{fig:defects1}(f). 

\begin{figure}[t]
\includegraphics[width=0.5\textwidth]{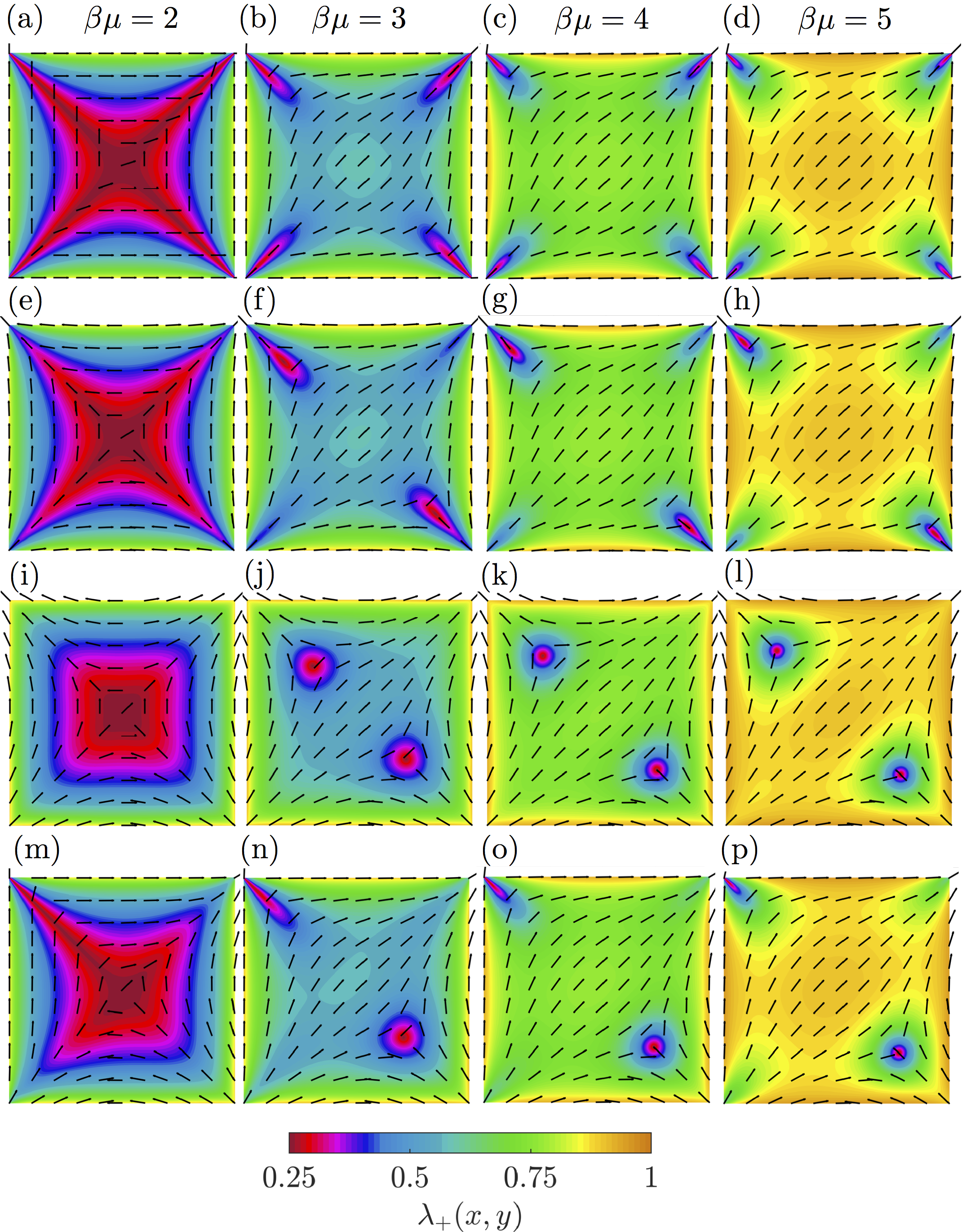}
\caption{Nematic director $\bf n$ (black lines) and the degree of two-dimensional ordering $\lambda_+(x,y)$ (colormap) for chemical potentials $\beta\mu=2, 3,4,5$ from left to right, for several anchoring parameter $\Delta$ (see Eq. \eqref{eq:directorboundary} and Fig. \ref{fig:sqgeometry}):(a)-(d) $\Delta=0$, (e)-(h) $\Delta=0.1$, (i)-(l) $\Delta=0.5$ and (m)-(p) $\Delta=0.5$ on $\Gamma_1$ and $\Gamma_4$, while $\Delta=0$ on $\Gamma_2$ and $\Gamma_3$. A two-dimensional isotropic state ($\lambda_+=1/4$) is colored in red. The cell size is the same as in Fig. \ref{fig:defects1}.}
\label{fig:defects2}
\end{figure}
\begin{figure}[t]
\includegraphics[width=0.45\textwidth]{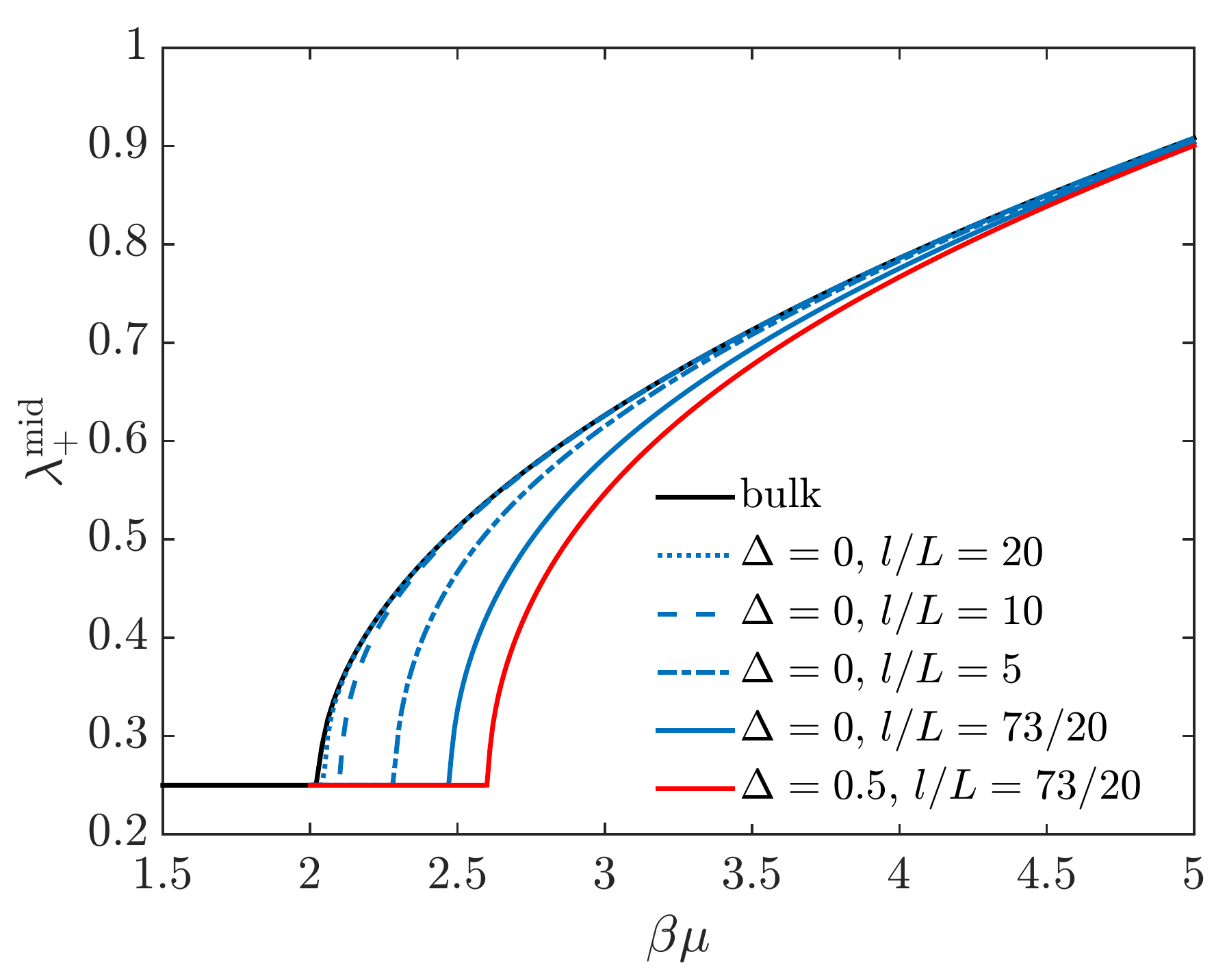}
\caption{The 2D order parameter evaluated at the center of the the cell $\lambda_+^\text{mid}=\lambda_+(0,0)$ as function of the chemical potential $\mu$ for various system sizes $l\times l$ and anchoring parameter $\Delta$ (dimensionless anchoring strength $w=10$). In black we show the 2D bulk phase transition, the 3D IN-transition occurs at $\beta\mu_{IN}=5.241$ beyond the scale of the plot.}
\label{fig:phasequasi2D}
\end{figure}
The type of possible structures are very sensitive to the boundary conditions that we impose. In Fig. \ref{fig:defects2} we show the various possibilities for different values of $\Delta$ as function of $\mu$. For $\Delta=0$ we see in Fig. \ref{fig:defects2}(a)-(d) that order is always reduced at the four corners, but these isotropic cores become smaller in size for larger $\mu$. The same happens when $\Delta=0.1$ in Fig. \ref{fig:defects2}(e)-(h), although now only in two of the four corners a 2D isotropic phase is found. In Fig. \ref{fig:defects2} (i)-(l) we set $\Delta=0.5$ and see that two defects are created at the center of the cell, which move apart along the diagonal when the total density is increased. When the center is perfectly ordered, $\lambda_+=1$, the defects reside at fixed positions, but not at the corners. A combination of the various structures can be found whenever $\Delta$ is different for the four boundaries. An example is shown in Fig. \ref{fig:defects2}(m)-(p) with $\Delta=0.5$ on $\Gamma_1$ and $\Gamma_4$, and $\Delta=0$ on $\Gamma_2$ and $\Gamma_3$.

A competition of the wall-induced ordering and the spontaneous nematic ordering always occurs beyond the chemical potential for which the center starts to order. Hence, we investigate $\lambda_+^\text{mid}\equiv\lambda_+(0,0)$ as function of $\mu$, shown for various system sizes $l$ and anchoring parameters $\Delta$ in Fig. \ref{fig:phasequasi2D}. We observe a second-order phase transition at $\mu>\mu_{IN}^{2D}$, with a shift that depends on the system size and on the nature of defects. In our calculations we also observe that a smaller anchoring strength $w$ shifts the phase transition closer to the 2D bulk one, the same as with increasing $l$. This can be understood from the observation that for the formation of a nematic phase in the center, it is necessary to counteract the wall-ordering effects. When the wall-ordering effects are stronger, which happens at larger $w$ or smaller $l$, a larger density and hence a larger $\mu$ is needed to spontaneously order the system.

Finally, we make some remarks on the location of the bulk phase transition of the quasi 2D setup that we have investigated here. Converting $\mu_{IN}^{2D}$ to a density, we find that the 2D bulk phase transition takes place at $c^*=1.95$, which should be compared with $\rho^*_{2D}=3\pi/2L^2$ or $c^*\approx 4.7$ for the two-dimensional Onsager model \cite{Kayser:1978}.  This significant discrepancy can be understood from the fact that we have not properly included the walls that confine the rods in the $xy$-plane, we simply set $Q_{zz}=-1/2$ and $Q_{zx}=Q_{zy}=0$. The Landau theory that we use is, however, effectively three-dimensional, and the excluded volume interactions in two dimensions are of a  different nature than in three dimensions. A better construction of the LdG theory would include these walls within a three-dimensional calculation. Another alternative is to use a two-dimensional Landau theory where the coefficients are to be determined from the bifurcation diagram of a two-dimensional Onsager theory. We expect that the qualitative features presented in this case, however, will not change.

\section{Nematic droplet}
\label{sec:nem}
Nematic droplets, or tactoids, differ from droplets of an isotropic fluid because they are elongated rather than spherical in shape. This has been observed in experiments, where the defect structures in the nematic texture have been investigated \cite{Kaznacheev:2002, Tang:2006, Tang:2007, Lavrentovich:2013, Schoot:2015}, but also in simulations \cite{Cuetos:2008}, {\color{black} and within the Zwanzig model \cite{Bier:2007}}. The shape of a tactoid can be determined within continuum theory \cite{Prinsen:2003} by minimizing the combined elastic and surface free energy
\begin{equation}
F=F_e+\int_{\partial V_\text{drop}} dS [\gamma_\parallel+(\gamma_\perp-\gamma_\parallel)(\hat{\boldsymbol{\nu}}\cdot{\bf n})^2], \label{eq:paul}
\end{equation}
at a finite and given volume of the droplet $V_\text{drop}$. Recall that the Frank elastic free energy $F_e$ is given in Eq. \eqref{eq:Frank}, and that $\gamma_\parallel$ $(\gamma_\perp)$ is the surface tension when the rods are aligned parallel (perpendicular) to the interface. Since the shape is an input parameter in the minimization procedure of Eq. \eqref{eq:paul}, one has to impose how the rods align along the interface of the droplet, and this is captured in the second term of Eq. \eqref{eq:paul}. Minimization of this free energy has been successful in determining the shapes of tactoids. The surface tensions that are used as fit parameters to match the theory with the experimentally observed shapes, however, disagree with the experimentally measured and theoretically obtained values of the surface tension \cite{Chen:1992, Koch:1999, Kaznacheev:2003, Shundyak:2003, Prinsen:2004, Shundyak:2006, Beek:2006, Puech:2010, Schoot:2015}. A hypothesis for this discrepancy is due to the simple choice of the Rapini-Papoular form \cite{Rapini:1969} to describe the anchoring at the IN interface, which is strictly speaking only true for a planar (non-curved) geometry. The question is how much curvature would renormalize the surface tensions of the flat geometry. Our aim is to calculate this within our Landau theory, because the surface effects of Eq. \eqref{eq:paul} should automatically be captured in our treatment. Such a calculation has been briefly touched upon for the thermotropic case in Ref. \cite{Menon:2008}, although only in a two-dimensional $xy$-geometry and with a focus on the time evolution rather than on the equilibrium tactoid sizes and shapes.
\begin{figure*}[t]
\includegraphics[width=0.95\textwidth]{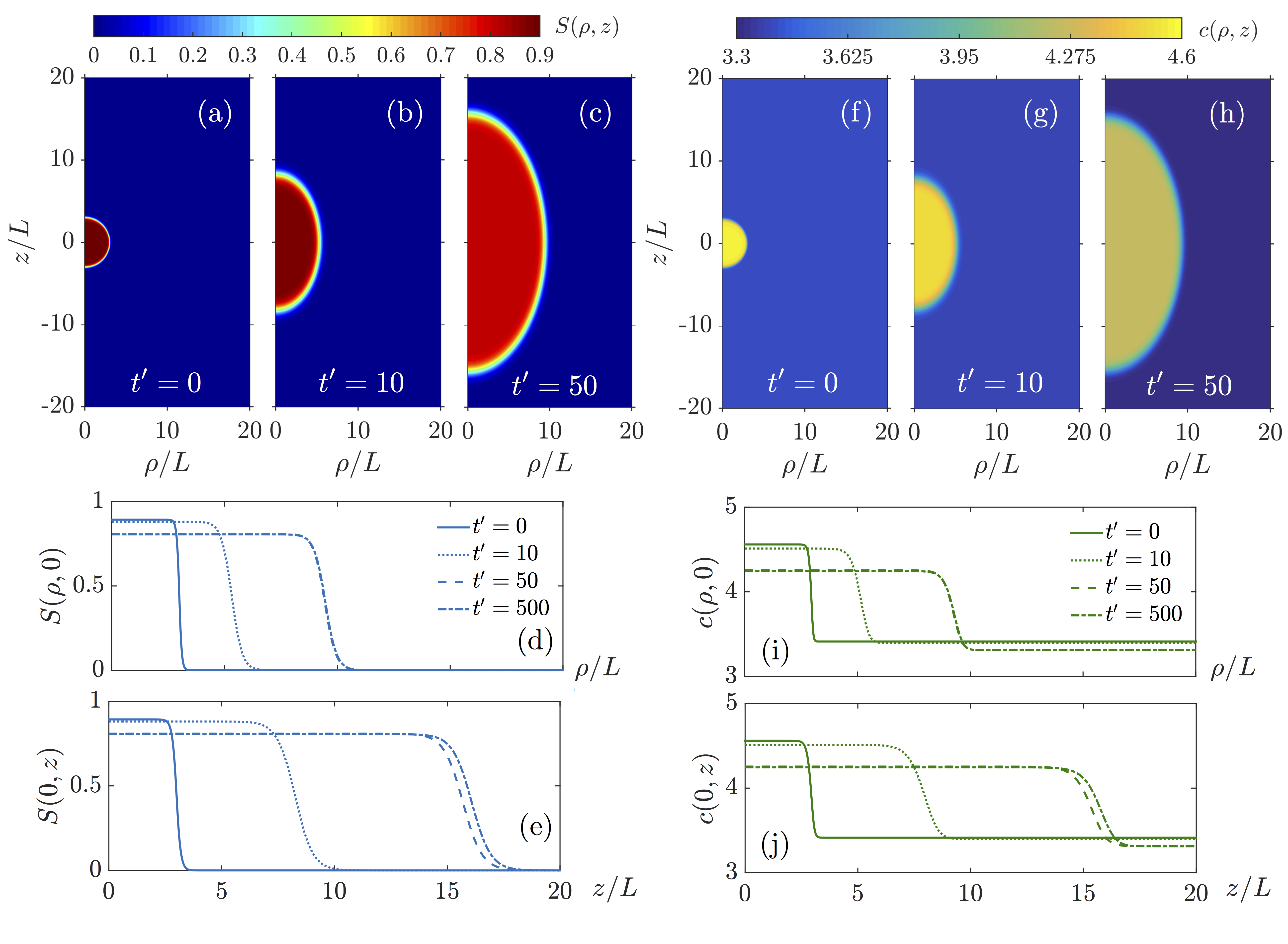}
\caption{Relaxation of a spherical liquid crystalline droplet of initial radius $r_0=3L$ prepared at $\beta\mu_0=5.5$ according to the dynamics of Eq. \eqref{eq:modelAmod} and constraint Eq. \eqref{eq:modelAconstraint}. In (a)-(c) we show snapshots of the order parameter profile $S(\rho,z)$ for various values of dimensionless time $t'$ showing that the spherical droplet becomes elongated over time. In (d) and (e) we show the order parameter profiles along the two symmetry axes of the droplet. In (f)-(h) we show the corresponding snapshots of the local dimensionless density $c(\rho,z)$ and the profiles along the symmetry axes in (i) and (j).}
\label{fig:tactoid1}
\end{figure*}

For simplicity we focus on tactoids with a homogeneous director field, ${\bf n}={\bf e}_z$, with ${\bf e}_z$ the unit vector in the $z$-direction. For such a director field all elasticity terms vanish, and hence the Landau grand potential is given by
\begin{align}
&\Delta\Omega[S]=\label{eq:tactoidom} \\
&\int d{\bf r}\Bigg[\frac{m_\parallel}{2\beta B_2}|\nabla S|^2+&\frac{m_\perp-m_\parallel}{2\beta B_2}({\bf n}\cdot\nabla S)^2 +\Delta\omega_b\Bigg] \nonumber.
\end{align}
It is easily seen that this free energy mimics the second term in Eq. \eqref{eq:paul}, with the identification that the surface normal $\hat{\boldsymbol{\nu}}$ is the same as $\nabla S$ on the surface defined by $S({\bf r}_*)=(1/2)\max_{{\bf r}\in V}S({\bf r})$, {\color{black} where we defined the IN interface to be the loci of points for which $S({\bf r})$ attains half its maximum value.} Thus, the square gradient term in Eq. \eqref{eq:tactoidom} is a surface tension contribution that results in droplets that tend to minimize their surface area, while the second term favours droplet shapes for which the misalignment between surface normal and director is minimal because $m_\perp>m_\parallel$. 

It turns out to be numerically difficult to find non-trivial solutions to the Euler-Lagrange equations of Eq. \eqref{eq:tactoidom} that are spatially inhomogeneous. For this reason we investigate instead a dynamical equation for $S$ \footnote{{\color{black}We neglect the noise terms in our calculations, since we are only interested in using Eq. \eqref{eq:modelA} as a pseudo dynamics to find stationary solutions to the Euler-Lagrange equation of Eq. \eqref{eq:tactoidom}. In Ref. \cite{Menon:2010} it is, however, listed how noise terms for a {\bf Q}-tensor theory can be constructed.}} assuming model-A (pseudo-)dynamics \cite{Hohenberg:1977},
\begin{equation}
\frac{\partial S}{\partial t}=-\Gamma\frac{\delta\Delta\Omega[S;\mu]}{\delta S({\bf r})}, \label{eq:modelA}
\end{equation}
and far-field boundary condition
\begin{equation}
\hat{\boldsymbol{\nu}}\cdot\nabla S({\bf r},t)=0, \quad {\bf r}\in \partial V. \label{eq:tactoidbound}
\end{equation}
As an initial condition we take a spherical droplet (or ``nucleus'')
\begin{equation}
S({\bf r},t=0)=\frac{3}{2}S_N^+(\mu)\Theta(r-r_0), \label{eq:tactoidinitial}
\end{equation}
where $\Theta$ is the Heaviside step function, $r_0$ is the radius of the initial droplet and $r=\sqrt{\rho^2+z^2}$, in terms of the cylindrical coordinates $(\rho,z)$.  Our goal is to look at a fixed $\mu$ for stationary solutions of Eq. \eqref{eq:modelA}, since they would also be solutions of the Euler-Lagrange equations. If we use Eq. \eqref{eq:modelA} for a fixed $r_0$, we only find solutions for which droplets keep growing if $\mu$ is too large, or droplets that keep shrinking if $\mu$ is too small. Consequently, there must be a critical chemical potential $\mu_c$ for which the droplet neither grows or shrinks when the optimal shape is attained. To find $\mu_c$ for a given shape and droplet volume we fix $\mu_0=\mu(t=0)$ and evaluate instead of Eq. \eqref{eq:modelA}, 
\begin{align}
&\frac{\partial S}{\partial t}= \label{eq:modelAmod} \\
&-\Gamma\frac{\delta}{\delta S({\bf r})}\left[\Delta\Omega[S;\mu_0]-\frac{\Delta\mu(t)}{\beta B_2}\int d{\bf r}\ c({\bf r},t;\mu_0+\Delta\mu(t))\right], \nonumber 
\end{align}
with a Lagrange multiplier $\Delta\mu(t)$ such that conservation of total number of particles
\begin{equation}
\int d{\bf r}\ c({\bf r},t;\mu_0+\Delta\mu(t))=\int d{\bf r}\ c({\bf r},t=0;\mu_0) \label{eq:modelAconstraint}
\end{equation}
is guaranteed. By definition $\Delta\mu(t=0)=0$. Working out the above Euler-Lagrange equation gives
\begin{align}
\frac{\beta B_2}{\Gamma}&\frac{\partial S}{\partial t}=m_\parallel\nabla^2 S+(m_\perp-m_\parallel)\frac{\partial^2S}{\partial z^2} \nonumber \\
&-2a\beta\{\mu^*-[\mu_0+\Delta\mu(t)]\}S+3bS^2-4dS^3. \label{eq:Eulertactoid}
\end{align}
Moreover, we find that we can approximate $\beta\mu(c_I)\approx A c_I+B$, with $A\approx2.25541$ and $B=-2.19865$, with a largest relative error of $0.6\%$ for $c_I\in[3,6]$. Using this approximation the constraint of Eq. \eqref{eq:modelAconstraint} can be rewritten as 
\begin{equation}
\int d{\bf r}\left[\frac{\beta\Delta\mu(t)}{A}+aS^2({\bf r},t)\right]=\int d{\bf r}\ aS^2({\bf r},t=0). \label{eq:approxcon}
\end{equation}

We solve Eq. \eqref{eq:Eulertactoid} and Eq. \eqref{eq:approxcon}, using the boundary condition Eq. \eqref{eq:tactoidbound} and initial condition Eq. \eqref{eq:tactoidinitial}  where we take for the system volume a cylinder of radius $20L$ and height $20L$, with the $xy$-plane as a symmetry plane for the droplet. When $r_0$ and $\mu_0$ are too small, the initial droplet disappears in time, but when they are chosen too big the droplet will eventually touch the system boundary and a ``planar'' interface along the radial direction will develop. However, here we will only consider the set of values for $\mu_0$ and $r_0$ that give stable droplets that do not change over time: this defines $\mu_c=\mu_0+\Delta\mu(t\rightarrow\infty)$.

In Fig. \ref{fig:tactoid1}(a)-(c) we show the typical relaxation of a spherical droplet as function of dimensionless time $t'=\Gamma L^3t/(\beta B_2)$ towards an elongated shape within our LdG theory, with the profiles along the two coordinate axes shown in Fig. \ref{fig:tactoid1}(d) and (e). Our LdG theory allows also for the calculation of $c(\rho,z)$ using Eq. \eqref{eq:densS}, which we show in Fig. \ref{fig:tactoid1}(f)-(h) and in Fig. \ref{fig:tactoid1}(i)-(j) along the two symmetry axes. The stationary solution that we find has a higher chemical potential than the bulk binodal, $\mu_c>\mu_{IN}$, which accounts for the Laplace pressure in this ensemble. 

Interestingly, the solution found for long times is very similar to the IN planar interface at coexistence. This is indicated by the interfacial shift $\delta$ and interfacial width $\tau$. We find for $S(\rho,0)$ and $c(\rho,0)$ that $\delta_\rho=0.219L$ and $\tau_\rho=0.576L$ to be compared with the flat plane result of  $\delta_\parallel=0.223L$ and $\tau_\parallel=0.586L$. Along the $z$ axis we find $\delta_z=0.372L$ and $\tau_z=0.978L$ to be compared with  $\delta_\perp=0.378L$ and $\tau_\perp=0.994L$. This shows that the surface tensions in both directions are effectively reduced for $\mu>\mu_{IN}$, since $\tau$ is smaller. However, from scaling arguments \cite{Schoot:1998}, we find that for $\gamma_\perp$ the relevant length scale is $L-(1/2)d^2/L$, with $d$ the possible displacement of the rods given the orientation distribution function. In contrast, for $\gamma_\parallel$ the relevant length scale is $d$. A higher $\mu$ is equivalent to a higher density and hence lower $d$. From the above scaling arguments we deduce that $\gamma_\perp$ should increase with $\mu$ while $\gamma_\parallel$ should decrease, which is not captured in our calculation.

\begin{figure}[t]
\includegraphics[width=0.5\textwidth]{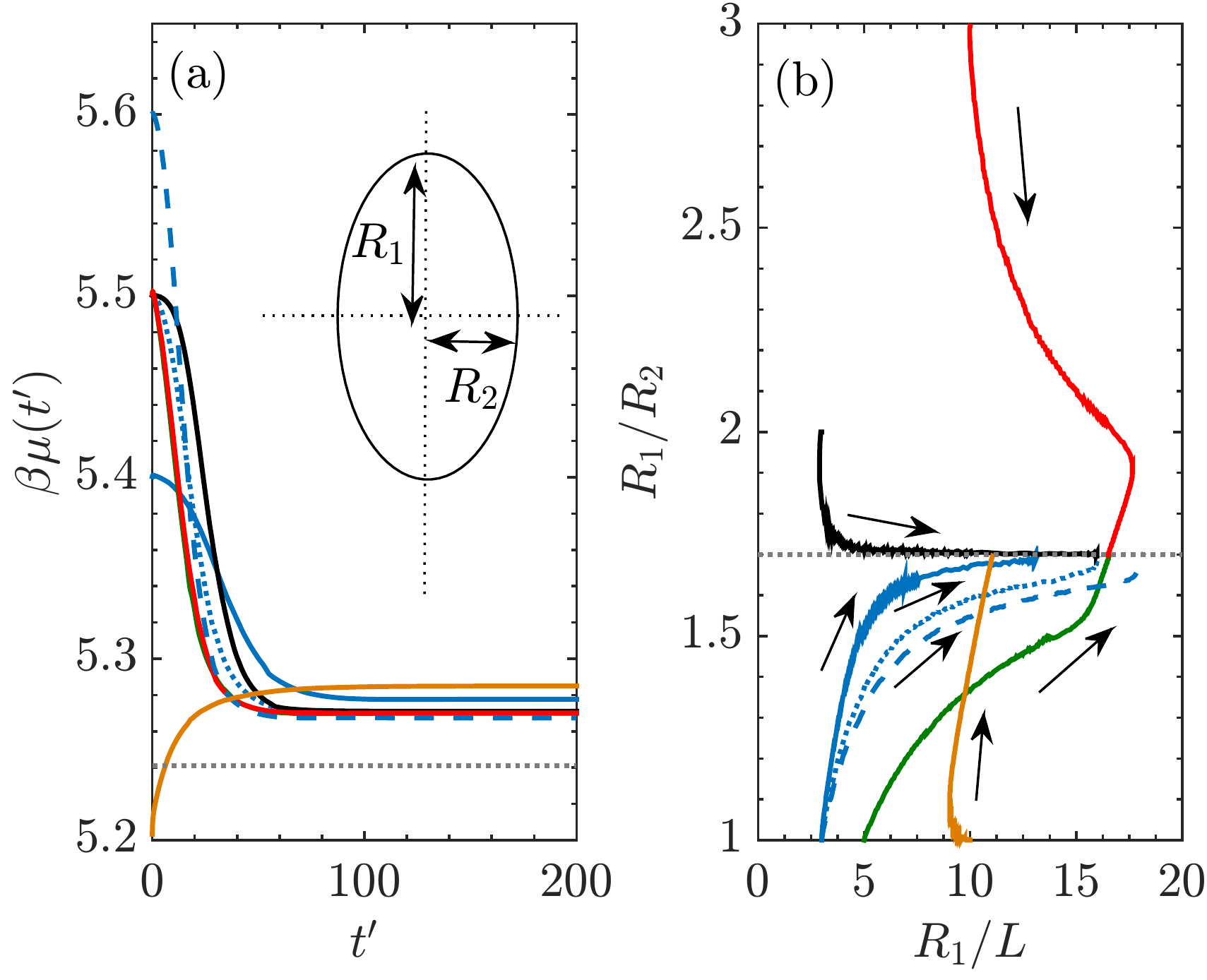}
\caption{(a) Time evolution of the chemical potential $\mu(t')=\mu_0+\Delta\mu(t')$ for various initial droplets and initial chemical potential $\mu_0$. We use the same color for initial droplets of the same $r_0$ and aspect ratio. In the inset we show the shape parameters $R_1$ and $R_2$ that we use to characterize the droplets. In (b) we indicate the time evolution of the aspect ratio $R_1/R_2$ as function of the droplet main axis $R_1$, where the various arrows indicate the flow of time. All lines end on the grey dotted line $R_1/R_2=1.7$ where the droplet does not change anymore, while the final droplet size $R_1(t'=\infty)$ depends crucially on $\mu_0$. The dotted blue lines correspond to the results shown in Fig. \ref{fig:tactoid1}.}
\label{fig:tactoid2}
\end{figure}

This latter observation is relevant when we investigate the aspect ratio $R_1/R_2$ of the stationary droplets, with $R_1$ the length of the main axis and $R_2$ the one of the minor axis, see the inset of Fig. \ref{fig:tactoid2}(a). It turns out that we always find $R_1/R_2=1.7$, which equals the ratio of the surface tensions determined from the planar geometry $\gamma_\perp/\gamma_\parallel=\sqrt{m_\perp/m_\parallel}$, see Eq. \eqref{eq:surftens}. Starting with different values of $\mu_0$ and $r_0$ (or even changing the aspect ratio of the initial droplet) we observe that $\mu_c$ only depends on $\mu_0$, see Fig. \ref{fig:tactoid2}(a). For small $\mu_c$, and hence lower values of the ``Laplace pressure", $\mu_c-\mu_{IN}$, we find larger droplets, as indicated by a larger $R_1$, although the differences are not large among the various final droplet sizes. However, regardless the final size, the aspect ratio is always $1.7$, which is illustrated by the various colored lines in Fig. \ref{fig:tactoid2}(b) converging to the dotted grey line. This is only to be expected when curvature effects are not important, which is also found within the macroscopic theory of Eq. \eqref{eq:paul} for homogeneous director fields with $\gamma_\perp$ and $\gamma_\parallel$ constant. However, we expect that $\gamma_\perp$ and $\gamma_\parallel$ depend on curvature. Moreover, in experiments \cite{Kaznacheev:2002, Tang:2006, Tang:2007, Lavrentovich:2013, Schoot:2015}, it is concluded that (i) the aspect ratio depends on $R_1$, (ii) the aspect ratio can become as large as $4$ to $5$ and (iii) the experimentally observed shapes have cusps at the endpoints on the axis parallel to the director field when the aspect ratio is larger than $2$. Only very large droplets are expected to have an aspect ratio of $1.7$. 

{\color{black} Interestingly, in simulations $R_1/R_2\sim 1.7$ is also found \cite{Cuetos:2008, Schilling:2009}, in contrast to the experiments where a larger $R_1/R_2$ up to $\sim 5$ has been observed \cite{Schoot:2015}. We note, however, that the rods in these simulations have a smaller particle aspect ratio ($L/D\sim20$) than in typical experiments ($L/D\sim 1000$). Moreover, rod flexibility, polydispersity and residual Van der Waals forces may be at play, and investigating these effects will be left for future work.  In the remainder of the text, we will instead speculate how the present LdG theory can produce $R_1/R_2> 1.7$. Whether or not these large tactoid aspect ratios can actually be found within Onsager theory remains, however, an open question.}

Clearly, increasing the aspect ratio is only possible within our theory by tuning the ratio $m_\perp/m_\parallel$. Indeed, we find in our calculations that we can obtain any desired aspect ratio by varying this quantity. Moreover, to make $R_1/R_2$ depend on the droplet size and hence on $\mu_c-\mu_{IN}$, the coefficients $m_\perp$ and $m_\parallel$ need to depend on $\mu$. This does not come as a surprise, since $l_1$ and $l_s$ are in general $\mu$ dependent. However, since $m_\perp=(2/3)(l_1+4l_s/3)$ and $m_\parallel=(2/3)(l_1+l_s/3)$ we find that the maximal aspect ratio that can be achieved by tuning $l_1$ and $l_s$ is $R_1/R_2=2$, while keeping $l_1,l_s>0$. However, we have to go to extreme limits to achieve this behaviour, $l_1$ should be close to zero or $l_s$ very large.

Another possibility to achieve a higher aspect ratio is by including higher order terms in the Landau expansion. The next order term has two derivatives and is third order in $\bf Q$. There are many symmetry-allowed terms that satisfy this condition, however, the only one that does not generate any new elasticity contributions and hence the only one that is quadratic in ${\bf n}$ is the term 
\begin{align}
\frac{2}{9}l_4Q_{\alpha\beta}\partial_\alpha Q_{\rho\sigma}\partial_\beta Q_{\rho\sigma}&=\frac{l_4}{2}S({\bf r})\left[({\bf n}\cdot\nabla S)^2-\frac{1}{3}|\nabla S|^2\right] \nonumber \\
&+\text{elasticity terms}, \label{eq:extraterm}
\end{align}
see Ref. \cite{Schiele:1983}, which holds for the uniaxial case. Another motivation why such a term is needed, is that it lifts the degeneracy on $K_{11}$ and $K_{33}$, which is also found within Onsager theory \cite{Straley:1973}, but not in LdG theory if the square gradient terms are only quadratic in ${\bf Q}$, see section II below Eq. \eqref{eq:Frank}. 

Clearly, including Eq. \eqref{eq:extraterm} increases $\gamma_\perp$, which is reflected by the positive sign of the first term. In contrast $\gamma_\parallel$ is reduced, since the second term is negative. Consequently, for $l_4>0$, the aspect ratio $R_1/R_2$ is increased. To reproduce the flat plane result for large droplets we expand $l_4=l_4^0(\mu-\mu_{IN})+\mathcal{O}[(\mu-\mu_{IN})^2]$ with $l_4^0>0$, and tried to do the calculation with this contribution \footnote{In principle the full $\mu$-dependence can be assessed by throughly comparing the Landau coefficients with the results of the elastic constants from Onsager theory \cite{Poniewierski:1979, Stecki:1980}.}. By construction larger droplets will have an aspect ratio closer to $R_1/R_2=1.7$, since $\mu_c$ will be closer to $\mu_{IN}$. While it was rather straightforward to get $R_1/R_2\sim2$ by tuning $l_4^0$, we found numerical difficulties when we tried to find a larger aspect ratio, because adding Eq. \eqref{eq:extraterm} introduces extra non-linearities in the Euler-Lagrange equation. Moreover, within all of our calculations, no cusps {\color{black}in the droplet shape} were found, while they are always experimentally observed for $R_1/R_2\gtrsim2$, {\color{black}see for example Ref. \cite{Schoot:2015}}. We hypothesize that these cusps are essential to have a large aspect ratio, and we speculate that we cannot find such solutions due to the adopted square-gradient approximation. Another possibility would be that model-A dynamics is not suitable to find these cusp(-like) solutions. Finally, we neglect any bipolarness in the director-field texture that may also be important \cite{Prinsen:2004, Schoot:2015}. It would be interesting to use the full $\bf Q$-tensor theory to capture this effect, since it is known within the macroscopic theory of Eq. \eqref{eq:paul} that $R_1/R_2$ depends on $R_1$ when bipolarness is included, even when $\gamma_\perp$ and $\gamma_\parallel$ are taken to be constant.

Despite its shortcomings here and there, our theory does show that adding a $\mu$-dependent elasticity term to the free energy allows us to predict an aspect ratio of $R_1/R_2=1.7-2$ that depends on the droplet size. The larger the droplet, the smaller $R_1/R_2$ since the ``Laplace pressure'' is smaller. If the elastic coefficients are assumed to be constant, we find no curvature effects on $R_1/R_2$. An improvement of the theory to obtain cusp-like solutions is needed, however. We expect that the cusps have a strong renormalizing effect on the surface tensions so that experimentally observed aspect ratios around $4$ can be achieved. The existence of these cusps is expected to be more important than that of the ``Laplace pressure", which turns out to be important whenever $R_1/R_2\sim2$.

 \section{Discussion and conclusions}
We have constructed a Landau theory for hard rods by a suitable order parameter expansion of the grand potential in Sec. \ref{sec:ldg}. The coefficients are determined by fitting them to the bulk coexistence data and the surface tensions in a planar geometry. This is different from the approach of Ref. \cite{Wittmann:2014} where the Frank elastic coefficients from a fundamental measure theory \cite{Wittmann:2015} are used to determine the square gradient coefficients. We have compared our results with known properties of Onsager theory, such as the (bulk) bifurcation diagram in Sec. \ref{sec:bulk} and characteristic length scales of the IN interface in Sec. \ref{sec:flat}. 

The remainder of the paper was a demonstration of the resulting Landau theory in more complex situations. We gave examples that were investigated before within Landau theory, but not yet within a density-dependent one.  In Sec. \ref{sec:hedg} we showed that we can assess the isotropic core size of hedgehog defects in terms of the rod length $L$ and investigated how the isotropic core size depends on the bulk density. Sufficiently far from the spinodal, we found a core size that is on the order of the length of the constituent particles, in contrast to a thermotropic hedgehog defect, which has a core size that is much larger than the length of the constituent molecules. In Sec. \ref{sec:rods} we studied the evolution of topological defects as function of density in confined quasi two-dimensional geometries and examined the effects of various boundary conditions. Sec. \ref{sec:nem} describes a novel application to find the curvature dependence of the surface tension in a self-consistent manner to explain the discrepancy between measurements and theoretical predictions of the surface tension for tactoids. Our calculations showed that the ``Laplace pressure'' renormalizes the surface tensions of the flat geometry for the perpendicular and parallel anchoring conditions, provided that a higher order $\mu$-dependent elastic coefficient is included. However, cusps in the equilibrium shapes of the tactoids are not found, and are expected to be important for the renormalization of the surface tensions.

We remark that the construction of the LdG theory is completely general. There are various ways of determining the coefficients, for which we have only shown one example. Moreover, these coefficients do not necessarily need to be fitted to Onsager theory, but can also be determined from comparisons with other theories for (hard) rods, such as fundamental measure theory \cite{Wittmann:2014}, or experiments and simulations. An interesting application would be to determine the Landau coeffcients from Khokhlov-Semenov theory to describe semi-flexible chains \cite{Khokhlov:1981, Khokhlov:1982, Vroege:1992}. Furthermore, we only studied the grand-canonical ensemble, however, for bulk systems it is also possible to consider the Gibbs free energy, where the pressure is the relevant intensive variable to be tuned to describe the density-dependent IN transition for hard rods. Ultimately, the final application determines which method is optimal. Finally, we hope that our findings will help to provide (qualitative) insights into problems that can be very hard to tackle within (inhomogeneous) Onsager theory or extensions thereof.

We acknowledge financial support of a Netherlands Organisation for Scientific Research (NWO) VICI grant funded by the Dutch Ministry of Education, Culture and Science (OCW) and from the European Union's Horizon 2020 programme under the Marie Sk\l{}odowska-Curie grant agreement No. 656327. This work is part of the D-ITP consortium, a program of the Netherlands Organisation for Scientific Research (NWO) funded by the Dutch Ministry of Education, Culture and Science (OCW).
\begin{widetext}
\section*{Appendix A: Euler-Lagrange equations for rods in square confinement}
In this appendix we derive the Euler-Lagrange equations that we used in section VI. For this we have to minimize $\Delta\Omega$ with respect to $\bf Q$ while taking into account that $\bf Q$ is traceless and symmetric, and that the order occurs in the $(x,y)$ plane. Therefore, we introduce the Lagrange multipliers $\lambda^B$, $\kappa_\rho^B$ and $\xi^B$ to ensure these constraints in the bulk, and for the surface we introduce likewise $\lambda^S$, $\kappa_\rho^S$ and $\xi^S$.

We define
\begin{align}
\beta B_2& \Delta\tilde\Omega[{\bf Q}]=\beta B_2\Delta\Omega[{\bf Q}]-\lambda^B\int_V d{\bf r}\delta_{\alpha\beta} Q_{\beta\alpha}({\bf r})-\frac{\kappa^B_{\alpha}}{2}\epsilon_{\alpha\beta\rho}\int_V d{\bf r}[Q_{\beta\rho}({\bf r})-Q_{\rho\beta}({\bf r})]-\xi^B\int_V d{\bf r} \delta_{\alpha z}\delta_{\beta z} \left(Q_{\alpha\beta}({\bf r})+\frac{1}{2}\right) \nonumber \\
&-\lambda^SL\int_{\partial V} dS \ \delta_{\alpha\beta}Q_{\beta\alpha}({\bf r})-\frac{\kappa^S_{\alpha}L}{2}\epsilon_{\alpha\beta\rho}\int_{\partial V} dS[Q_{\beta\rho}({\bf r})-Q_{\rho\beta}({\bf r})]-\xi^SL\int_{\partial V} dS \delta_{\alpha z}\delta_{\beta z} \left(Q_{\alpha\beta}({\bf r})+\frac{1}{2}\right).
\end{align}
The Lagrange multipliers $\lambda^B$ and $\lambda^S$ ensure that $\bf Q$ is traceless in the bulk and surface respectively, while $\kappa_\alpha^B$ and $\kappa_\alpha^S$ ($\alpha=1,2,3$) ensure that $\bf Q$ is symmetric in the bulk and on the surface respectively. It is then straightforward to find the Euler-Lagrange equations by setting $\delta\Delta\tilde{\Omega}/\delta Q_{\alpha\beta}({\bf r})=0$ to find for $\alpha,\beta=1,2,3$ and ${\bf r}\in V$
\begin{equation}
l_1\partial_\lambda^2 Q_{\alpha\beta}+l_2\partial_\lambda\partial_\alpha Q_{\lambda\beta}+l_3\partial_\lambda\partial_\beta Q_{\alpha\lambda}-3a\beta(\mu^*-\mu)Q_{\beta\alpha}+9b\ Q_{\beta\rho}Q_{\rho\alpha}-4d Q_{\lambda\rho}Q_{\rho\lambda}Q_{\beta\alpha}=\lambda^B\delta_{\alpha\beta}+\kappa^B_\rho \epsilon_{\rho\alpha\beta}+\xi^B\delta_{\alpha z}\delta_{\beta z},  \label{eq:bulk}
\end{equation}
while for ${\bf r} \in\partial V$
\begin{equation}
\frac{4}{9}[l_1\partial_\lambda Q_{\alpha\beta}\hat{\nu}_\lambda+l_2\partial_\lambda Q_{\lambda\beta}\hat{\nu}_\alpha+l_3\partial_\beta Q_{\alpha\lambda}\hat{\nu}_\lambda]+w(Q_{\alpha\beta}-Q^0_{\alpha\beta})=\lambda^S\delta_{\alpha\beta}+\kappa^S_\rho\epsilon_{\rho\alpha\beta}+\xi^S\delta_{\alpha z}\delta_{\beta z},
\label{eq:bound}
\end{equation}
with $\hat{\boldsymbol{\nu}}$ being the unit surface normal. Setting the derivatives with respect to the Lagrange multipliers to zero, gives the constraints
\begin{align}
\text{Tr}({\bf Q})=0, \quad Q_{\alpha\beta}=Q_{\beta\alpha}\quad (\alpha,\beta=x,y,z), \quad Q_{zz}=-\frac{1}{2}. \label{eq:constraints}
\end{align}
The Lagrange multipliers can be determined by evaluating the $zz$ component of Eq. \eqref{eq:bulk},
\begin{equation}
\lambda^B+\xi^B=\frac{3}{2}a\beta(\mu^*-\mu)+\frac{9}{4}b+2d Q_{\lambda\rho}Q_{\rho\lambda},
\end{equation}
while taking the trace of Eq. \eqref{eq:bulk} gives
\begin{equation}
3\lambda^B+\xi^B= 2l_s\partial_\rho\partial_\lambda Q_{\lambda\rho}+9b Q_{\lambda\rho}Q_{\rho\lambda},
\end{equation}
hence
\begin{equation}
\lambda^B=l_s\partial_\lambda\partial_\rho Q_{\rho\lambda}+\frac{9}{2}b\left(Q_{\lambda\rho}Q_{\rho\lambda}-\frac{1}{4}\right)-\frac{3}{4}a\beta(\mu^*-\mu)-d Q_{\lambda\rho}Q_{\rho\lambda}. \label{eq:lambda}
\end{equation}
The effect of the Lagrange multipliers $\kappa_\rho$ is to symmetrize Eq. \eqref{eq:bulk} over the indices $\alpha$ and $\beta$. A similar calculation can be performed to determine the surface Lagrange multipliers. Combining Eq. \eqref{eq:bulk}-\eqref{eq:constraints} and Eq. \eqref{eq:lambda} results in the Euler-Lagrange equations Eq. \eqref{eq:EL1} and boundary condition Eq. \eqref{eq:EL2}.
\end{widetext}

\bibliographystyle{apsrev4-1} 
\bibliography{literature1} 

\end{document}